\documentclass[12pt]{article}

\usepackage[utf8]{inputenc}
\usepackage[top=.55in, bottom=.55in, left=0.55in, right=0.55in, headheight=13pt, footskip=0.4in]{geometry}
\usepackage{mathpazo}
\usepackage[T1]{fontenc}
\usepackage{graphicx}
\usepackage{amsmath,amssymb,amsfonts}
\usepackage{physics2}
\usephysicsmodule{ab.legacy}
\usepackage{blkarray}  
\usepackage{mathtools} 
\usepackage{booktabs}
\usepackage{newunicodechar}
\newunicodechar{ȩ}{\c{e}}
\usepackage{float}
\usepackage{caption}
\usepackage{xcolor}
\usepackage{authblk}
\usepackage{setspace}
\usepackage{tabularx}
\usepackage{xltabular}
\newcolumntype{L}{>{\raggedright\arraybackslash}X}
\usepackage{microtype}
\usepackage{fancyhdr}
\usepackage[backend=biber, style=numeric-comp, sorting=none, maxbibnames=10, giveninits=true]{biblatex}
\usepackage[hidelinks]{hyperref}
\usepackage{orcidlink} 
\usepackage{cleveref}
\usepackage[title]{appendix}

\graphicspath{{figures/}{./}}
\crefname{figure}{Figure}{Figures}
\crefname{table}{Table}{Tables}
\crefname{equation}{Eq.}{Eqs.}
\crefname{section}{Section}{Sections}
\crefname{appendix}{Appendix}{Appendices}
\newcommand{\kbar}{\bar{k}}

\begin{document}

\title{\textbf{Cutting Corners or Saving Lives? \\ The Evolutionary Stability of Ethics in Pharma Medicine}}

\author[1]{Alanna Polyak\,\orcidlink{0009-0001-2373-3659}}
\author[1]{Thomas Zdyrski\,\orcidlink{0000-0003-3039-172X} \thanks{Corresponding Author. Email: Thomas.Zdyrski@dartmouth.edu}}
\affil[1]{Department of Mathematics, Dartmouth College, Hanover, NH 03755, United States of America}

\maketitle

\begin{abstract}
The pharmaceutical industry operates at the intersection of financial incentives, state oversight, and public scrutiny. We model corporate compliance with a three-population replicator system of ethical firms, active regulators, and active public scrutiny, whose payoffs are motivated by a prisoner's dilemma of first-mover advantage, an inspection game between firms and the FDA, and a volunteer's dilemma in early-stage research. The eight payoff constants are fitted by bounded L-BFGS-B optimization to composite indices for 2010--2024, and a fully sourced dataset of fifteen heavily penalized firms links the model's strategy types to R\&D intensity and penalties. The calibrated system has no interior fixed point: an exact identity shows that public scrutiny decays at a rate set by the difference between the regulators' and the public's cost-to-benefit ratios. Without public scrutiny the firm--regulator dynamics are conformally Hamiltonian, so the two populations cycle neutrally around a center, and Lyapunov exponents give no evidence of chaos. Bureaucratic friction $k$, which enters compliance costs as $k^3$, organizes two thresholds: above $k^*\approx0.75$ regulators saturate and compliance rests on public scrutiny, and above $k^{**}\approx0.81$ compliance collapses. The empirical friction index crosses both between 2016 and 2020, as the compliance index falls from $0.85$ to $0.60$. Relative to revenue, penalties are highest for the firms with the lowest R\&D intensity. Ethical compliance is therefore an oscillating rather than a permanent state, persisting only while friction stays below the combined incentive of subsidies, fines, and public penalties.

\end{abstract}

\section{Introduction}
Evolutionary game theory (EGT) provides a mathematical framework for strategic interactions between boundedly rational agents over extended periods. Originally formulated by John Maynard Smith to model competition in biological populations \cite{smith1982}, EGT has since been applied widely to socio-economic behavior in oligopolistic markets \cite{weibull1995}. Formally, EGT replaces the perfect rationality of classical game theory with replicator dynamics, in which the growth rate of a strategy is proportional to its fitness advantage over the population average \cite{taylor1978}; the three canonical games invoked below (the prisoner's dilemma, the inspection game, and the volunteer's dilemma in \cref{sec:games}) each motivate one term of the fitness functions of \cref{sec:model}. Closely related three-population replicator models have recently been applied to pharmaceutical regulation itself, most notably Feng et al.'s evolutionary game of regulators, drug manufacturers, and patients in the context of accessible drug packaging in China \cite{feng2024}, which we return to in the Discussion.
Evolutionary game treatments of neighboring problems include
reputation-based compliance mechanisms \cite{tavoni2012survival}
and divisions of labor on networks \cite{nirjhor2026evolution},
delayed interactions within replicator dynamics \cite{fic2025compartment},
and
public goods games that incorporate anti-corruption strategies \cite{fang2020singaporean}.
In the pharmaceutical industry, establishing a sustainable culture of ethics is complicated by these interlocking games.

The modern pharmaceutical sector operates under a persistent tension: companies are legally bound by fiduciary duty to maximize shareholder returns, yet they are simultaneously tasked with developing life-saving public health interventions. Bringing a single novel drug to market costs an estimated \$1.4 billion in out-of-pocket spending, or about \$2.6 billion once the cost of capital is included (2013 dollars) \cite{dimasi2016}, and development takes roughly a decade \cite{cbo2021}. These large upfront costs create an environment in which the temptation to bypass rigorous safety standards, accelerate trial phases, or inflate prices is strong. Our central research question asks: \textit{Why do pharmaceutical companies repeatedly make unethical decisions (e.g., price gouging, off-label marketing, data concealment) even if it damages them in the long run? Can regulatory pressure and public monitoring shift the system toward ethical behavior?}

This question is timely given recent events in the pharmaceutical industry. Our compiled compliance index falls from its 2016 peak of $0.85$ to $0.60$ in 2020 and $0.25$ by 2024, while the bureaucratic-friction index rises monotonically from $0.60$ in 2010 to $1.05$ in 2024, reaching $0.88$ in 2020 (\cref{sec:params}). \Cref{sec:jac} shows that the calibrated model has two friction thresholds, $k^*\approx0.75$ and $k^{**}\approx0.81$, and the friction index passes both between 2016 and 2020. A goal of this paper is to ask whether a mechanistic model of competing populations can reproduce that decline, and if so, what it implies about which policy levers could arrest it.

To address this, we model three intersecting games played among three strategic populations: \textbf{ethical firms ($x$)}, \textbf{active regulators ($y$)}, and \textbf{active public scrutiny ($z$)}.

First, the \textbf{prisoner's dilemma} describes firm-versus-firm competition. Since the first company to patent and launch a novel therapy can secure a large market share, firms face strong pressure to cut costs, rush trials, and bypass rigorous safety standards. When a rare defecting firm appears in a population of cooperators, it gains a fitness advantage, driving the population toward a Pareto-inferior Nash equilibrium \cite{axelrod1984}.

Second, the \textbf{inspection game} frames the relationship between firms and regulators through hawk--dove dynamics, a well-documented structure in regulatory environments \cite{avenhaus2002}. Regulatory bodies such as the FDA operate with finite resources and cannot audit every trial. Darrow et al.\ map the historical progression of FDA regulation, showing how changes in state oversight alter corporate strategies \cite{darrow2020}. When oversight relaxes, the expected payoff of defecting increases and cost-cutting spreads; when regulators respond to a crisis by over-enforcing, innovation is burdened by compliance costs.

Third, the \textbf{volunteer's dilemma} governs early-stage R\&D. Foundational science is expensive and prone to failure. Returns to drug R\&D are highly skewed, with a small share of products earning most of the returns \cite{grabowski2002}, and Comanor and Scherer argue that mergers among large firms reduce the number of parallel research efforts and with it overall innovation \cite{comanor2013}. A rational firm therefore has an incentive to let rivals absorb the cost and risk of early research before building on the results, which creates a systemic innovation deficit known as the ``valley of death'': the gap between publicly funded basic discovery and commercially funded late-stage development, in which promising candidates stall because no actor is willing to bear the cost of the intermediate, high-failure-rate stage.

Current literature establishes the foundation for these interactions but often falls short of continuous, dynamic modeling. Grepperud and Pedersen modeled the positioning game that precedes pharmaceutical price negotiations as a static game between a purchaser and a firm \cite{grepperud2020}, while Jarrell and Peltzman quantified the shareholder-wealth losses associated with product recalls \cite{jarrell1985}. Our primary contribution is synthesizing these isolated concepts into a single continuous-time evolutionary framework calibrated against a real-world dataset, and characterizing its long-run behavior analytically.

\section{Methods}\label{sec:methods}
This section describes how the empirical data were gathered, defines the replicator system and the games that motivate its payoff terms, analyzes its fixed points and long-run behavior, and describes the numerical methods used for calibration and for the spatial, stochastic, and delayed extensions.

\subsection{Data Compilation and Firm Classification}\label{sec:data}
A major contribution of this project is a custom dataset (Appendix~\ref{app:firms}, \cref{tab:full_firms}). Since no single public database tracks game-theoretic metrics for pharmaceutical behavior, we manually compiled and aggregated this dataset. We sourced R\&D spending percentages from historical \textbf{SEC 10-K filings (2010--2024)} to determine the baseline internal investment of each corporation. Penalty severity and compliance-violation logs were compiled from the Good Jobs First Violation Tracker, FDA warning-letter dashboards, and Department of Justice settlement press releases. The penalty measure for every firm is its total federal and state financial penalties over 1991--2021 as tabulated by Public Citizen (Table~4 of \cite{publiccitizen2024}), and R\&D intensity is fiscal-2019 R\&D expense divided by fiscal-2019 revenue, both as reported in the firm's annual report or Form 10-K. To compare penalties with firm size we divide them by each firm's fiscal-2019 revenue, taken from its annual report or Form 10-K (Appendix~\ref{app:firms}).

To establish macro-economic validity, this regulatory data was cross-referenced with aggregate industry analyses, including the Public Citizen report \cite{publiccitizen2024}. That report tabulated 482 federal and state settlements with pharmaceutical manufacturers between 1991 and 2021, totaling \$62.3 billion in financial penalties; four companies (Purdue, Johnson \& Johnson, GlaxoSmithKline, and Pfizer) accounted for 48\% of that total, and unlawful promotion was the violation category with the largest penalties (\$22.3 billion). This concentration mirrors the defection pattern modeled in our simulation and motivates the classification of 15 major firms into three strategy groups: S1 cooperators (ethical R\&D), S2 defectors (cost cutters), and mixed-strategy actors.

To guard against selective inclusion, the sampling frame and selection criteria were fixed prior. A firm is included if (i) it appears on Public Citizen's list of the twenty largest penalty totals for 1991--2021 (Table~4 of \cite{publiccitizen2024}), (ii) it still existed as an independent company in fiscal 2019, the reference year for revenue and R\&D, and (iii) all three quantities the analysis uses (R\&D expense, revenue, and the penalty total) are published for it. Fifteen of the twenty companies meet all three conditions. Of the other five, Schering-Plough (merged into Merck in 2009), TAP Pharmaceutical Products (dissolved in 2008), and Serono (acquired by Merck KGaA in 2007) had no 2019 accounts; Purdue Pharma is privately held and publishes no audited revenue or R\&D figures; and for Reckitt Benckiser, a consumer-goods company whose prescription-drug business was demerged as Indivior in 2014, we could not obtain a fiscal-2019 R\&D figure from its published accounts. Indivior itself is on the list and is included. Requiring every firm to have all three numbers means that every firm appears in every firm-level figure and table. Classification into the S1/Mixed/S2 zones then uses fixed R\&D thresholds ($>$21\%, 10--21\%, $<$10\% of revenue) and not the penalty data. To ensure reproducibility, we collected data using a fixed protocol. For each firm we (i) extracted annual R\&D-to-revenue ratios from SEC 10-K filings over 2010--2024; (ii) recorded every federal settlement, warning letter, and recall logged in the Good Jobs First Violation Tracker, the FDA warning-letter dashboard, and DOJ press releases over the same window; and (iii) cross-checked each entry against the aggregate totals in the Public Citizen report \cite{publiccitizen2024}. Throughout the paper a series means one raw indicator recorded once per observation year, such as the annual count of Class~I recalls, the number of FDA warning letters issued, the number of FDA inspections, or an annual media-sentiment score. Each raw series was normalized to $[0,1]$ and combined into the composite indices using the formulas of \cref{sec:params}; the coding scheme, source URLs, and access dates are recorded for every firm in the released dataset (\texttt{data/firm\_classification.csv}).

The resulting fifteen firms are listed in \cref{tab:full_firms}. The four S2 firms (Teva, Mylan, Indivior, and Abbott) spend less than $10\%$ of revenue on R\&D; three earn most of their revenue from generic or legacy products, and Abbott's prescription-drug business was separated as AbbVie in 2013, leaving a devices and diagnostics company. The seven Mixed firms (Pfizer, GlaxoSmithKline, Johnson \& Johnson, Novartis, Amgen, Allergan, and Mallinckrodt), with intensities between $10\%$ and $21\%$, are manufacturers that sustain substantial pipelines yet appear in the same settlement records, predominantly for off-label promotion; a three-zone classification requires this intermediate category of firms that are neither pure cooperators nor pure defectors. The four S1 firms (Merck, Eli Lilly, AstraZeneca, and Bristol Myers Squibb) spend more than $21\%$ of revenue on R\&D.

Both cut-offs fall in natural gaps of the data, and are further validated. The $10\%$ cut-off lies in the largest gap of the R\&D distribution, between Abbott ($7.6\%$) and Mallinckrodt ($11.0\%$): every firm below it earns its revenue mainly from generic, legacy, or non-drug products, and every firm above it runs a research pipeline of its own. The $21\%$ cut-off lies between Novartis ($19.8\%$) and Merck ($21.1\%$) and corresponds to the industry-wide R\&D intensity reported by the Congressional Budget Office, about one quarter of revenues measured net of rebates \cite{cbo2021}; because 10-K revenue is reported before rebates, that benchmark is equivalent to a somewhat lower share of reported revenue, so S1 firms are those investing at or above the industry norm. The counts per zone ($4$ S2, $7$ Mixed, $4$ S1) were not chosen: they are what the two cut-offs give for the fifteen firms the rule selects. The middle zone is the largest because most large research-based manufacturers spend between a tenth and a fifth of their revenue on R\&D. We did not force the zones to the same size, since that would mean choosing which firms to add or drop. Instead, Panel~18.3 shows every firm individually, and the ordering of the zone medians does not hinge on any single firm: dropping each firm in turn leaves the S2 median between $10.5\%$ and $13.8\%$ of a year's revenue, the Mixed median between $7.7\%$ and $9.8\%$, and the S1 median between $4.2\%$ and $7.0\%$, so the three ranges never overlap. Since the zone thresholds were fixed before penalties were examined, a firm's classification is a function of its R\&D intensity alone; the penalty column is therefore an outcome variable rather than a criterion, which is what permits the relationship in \cref{fig:rnd_penalty} to be read as a true finding. The sample is deliberately stratified rather than random, so it supports comparison across strategy types but not inference about the industry's unconditional distribution, a limitation revisited in \cref{sec:limitations}.

\subsection{Fitness Functions and Replicator Dynamics}\label{sec:model}
Following Hofbauer and Sigmund \cite{hofbauer1998}, we assume three infinite, well-mixed populations, each choosing between two actions. Evolutionary game theory is applied to all three populations because each faces a selection pressure that rewards a successful strategy with a larger share of its own population: firms compete for capital and market share, regulators' institutional survival and budget depend on demonstrably successful oversight, and members of the public derive social benefit from scrutinizing unethical firms and withdraw that effort when it is unrewarded. Let $x(t)\in[0,1]$ be the proportion of ethical (complying) firms, $y(t)\in[0,1]$ the proportion of active (inspecting) regulators, and $z(t)\in[0,1]$ the proportion of the public actively scrutinizing firms. The state space is therefore the unit cube $[0,1]^3$.

Writing $f_1$ for the firm payoff, $f_2$ for the regulator payoff, and $f_3$ for the public payoff, the fitness advantage of an ethical firm ($E$) over a defecting firm ($D$) is
\begin{equation}
f_1^E - f_1^D = -(C_{base} + k^3 y) + y(S_{pol} + \alpha) + z\gamma ,
\label{eq:fE}
\end{equation}
and the advantages of active over passive regulators and public are
\begin{equation}
f_2^{\text{active}} - f_2^{\text{passive}} = (1-x)B_r - C_r,
\qquad
f_3^{\text{active}} - f_3^{\text{passive}} = (1-x)B_p - C_p .
\label{eq:payoffs23}
\end{equation}
Here $k$ is bureaucratic friction (``red tape''), which raises the cost of compliance as $k^3$. The cubic is a modeling choice representing costs that grow faster than linearly in regulatory complexity, in the spirit of ``Eroom's law'' for declining R\&D efficiency \cite{scannell2012}; it is a power law, not an exponential. $C_{base}$ is the baseline cost of compliance, $S_{pol}$ the FDA subsidy to complying firms, $\alpha$ the FDA fine levied on defectors, and $\gamma$ the public penalty, which models the shareholder-wealth losses documented by Jarrell and Peltzman \cite{jarrell1985}. Regulators ($y$) and the public ($z$) pay overhead costs $C_r$ and $C_p$ to monitor the market and gain institutional and social benefits $B_r$ and $B_p$ when they catch a defecting firm, which happens at rate $(1-x)$. \Cref{sec:games} derives \cref{eq:fE,eq:payoffs23} from explicit payoff matrices.

Collecting the three population shares as $x_i \coloneqq (x,y,z)$ for $i=1,2,3$ and applying the standard replicator equation $\dot{x}_i = x_i\,(f_i - \bar{f}_i)$, in which $\bar{f}_i$ denotes the average payoff within population $i$ (the average over its two actions weighted by their current shares, not an average across the three populations), gives the continuous-time system
\begin{align}
\dot{x} &= x(1-x) \left[ -(C_{base} + k^3 y) + y(S_{pol} + \alpha) + z\gamma \right], \label{eq:x} \\
\dot{y} &= y(1-y) \left[ (1-x)B_r - C_r \right], \label{eq:y} \\
\dot{z} &= z(1-z) \left[ (1-x)B_p - C_p \right]. \label{eq:z}
\end{align}
For the analysis below it is convenient to write \crefrange{eq:x}{eq:z} with generic coefficients,
\begin{equation}
\dot x = x(1-x)\,\underbrace{(a_0 + a_y y + a_z z)}_{\textstyle F},\qquad
\dot y = y(1-y)\,\underbrace{(b_0 + b_x x)}_{\textstyle G},\qquad
\dot z = z(1-z)\,\underbrace{(c_0 + c_x x)}_{\textstyle H},
\label{eq:generic}
\end{equation}
with $a_0\coloneqq-C_{base}$, $a_y\coloneqq S_{pol}+\alpha-k^3$, $a_z\coloneqq\gamma$, $b_0\coloneqq B_r-C_r$, $b_x\coloneqq-B_r$, $c_0\coloneqq B_p-C_p$, and $c_x\coloneqq-B_p$. Only $a_y$ depends on $k$.

The eight benefit and cost constants ($C_{base}$, $S_{pol}$, $\alpha$, $\gamma$, $B_r$, $C_r$, $B_p$, $C_p$) are obtained by fitting the system to the empirical indices with the L-BFGS-B calibration of \cref{sec:ml}; their values and bounds are listed in \cref{tab:params} (Appendix~\ref{app:params}). The robustness of the results to $\pm 20\%$ perturbations in each constant is reported in the sensitivity analysis of Panel~4.4 (\cref{fig:policy}).

\subsection{Game-Theoretic Foundations of the Payoff Terms}\label{sec:games}
The model combines three games into a single continuous phase space.
\begin{enumerate}
    \item \textbf{The asymmetric inspection game (hawk--dove):} The interaction between firms and the FDA turns on the inspection probability $y$. An active regulator imposes the friction cost $k^3$ on a complying firm, grants it the subsidy $S_{pol}$, and fines a defector $\alpha$; a passive regulator does none of these. In normal form (row player's payoff first),
\begin{equation}
\begin{blockarray}{ccc}
 & \text{inspect} & \text{not inspect} \\
\begin{block}{c(cc)}
 \text{comply} & R-C_{base}-k^3+S_{pol},\;-C_r & R-C_{base},\;0 \\
 \text{defect} & R-\alpha,\;B_r-C_r & R,\;0 \\
\end{block}
\end{blockarray}
\label{eq:payoffinsp}
\end{equation}
where $R$ is a firm's revenue. Against a regulator population that inspects with probability $y$, the expected payoff difference between complying and defecting is $y\,(S_{pol}+\alpha-k^3) - C_{base}$, which is \cref{eq:fE} with $z=0$; against a firm population that complies with probability $x$, the regulator's expected advantage from inspecting is $(1-x)B_r - C_r$, which is \cref{eq:payoffs23}. The public plays the same game against firms with $(\gamma, B_p, C_p)$ in place of $(\alpha, B_r, C_r)$ and no subsidy or friction term, contributing $z\gamma$ to \cref{eq:fE} and the second expression in \cref{eq:payoffs23}.
    \item \textbf{The multi-player prisoner's dilemma:} Firms also compete to be first to market. In two-player form,
\begin{equation}
\begin{blockarray}{ccc}
 & \text{comply} & \text{defect} \\
\begin{block}{c(cc)}
 \text{comply} & R-C_{base},\;R-C_{base} & R-C_{base}-\Delta,\;R \\
 \text{defect} & R,\;R-C_{base}-\Delta & R-\alpha,\;R-\alpha \\
\end{block}
\end{blockarray}
\label{eq:payoffpd}
\end{equation}
where $\Delta$ is the first-mover loss suffered by a complying firm whose rival reaches market first. This is a prisoner's dilemma whenever $0 < C_{base}<\alpha<C_{base}+\Delta$: defection is dominant although mutual compliance is better for both. The first-mover loss $\Delta$ is not separately identifiable from the composite indices, so in the calibrated model the content of this game is carried by the baseline cost $C_{base}>0$, which makes defection the better response in the absence of inspection or scrutiny when $C_{base} > \alpha$.
    \item \textbf{The volunteer's dilemma:} In early-stage R\&D, if $N$ firms exist, the cost $C_v$ to volunteer (pioneer a new drug class) yields a systemic benefit $B$; if no firm volunteers, all receive $0$. The utility of volunteering is $U_v = B - C_v$, while the expected utility of free-riding is
\begin{equation*}
U_{f} = B \left[1 - P(\text{no other firm volunteers})\right] = B\left[1 - (1-x)^{N-1}\right].
\end{equation*} Since this is an $N$-player game, it is stated in utility form rather than as a $2\times2$ matrix. It motivates the subsidy $S_{pol}$: a non-punitive transfer to the firms that bear the cost of foundational research.
\end{enumerate}

\subsection{Externally Calculated Parameters and Explicit Variables}\label{sec:params}
Several quantities are computed before, and independently of, the numerical integration of \crefrange{eq:x}{eq:z}: they are calculated once from the source data, held fixed while the ordinary differential equation (ODE) solver runs, and never updated by it. They therefore enter the model as data rather than as free parameters. Each raw annual series $v$ (\cref{sec:data}) is first placed on a common scale by min--max normalization over the 2010--2024 window,
\begin{equation}
\tilde{v}_t \;=\; \frac{v_t - \min_\tau v_\tau}{\max_\tau v_\tau - \min_\tau v_\tau} \;\in\; [0,1].
\label{eq:minmax}
\end{equation}
Throughout, a tilde denotes this normalized form of a raw series. Min--max scaling is used in preference to a $z$-score because the replicator state variables are population shares and must lie in $[0,1]$ for \crefrange{eq:x}{eq:z} to be well posed. The four composite indices are fixed linear combinations of the normalized series:
\begin{align}
\text{Ethical firms:}\quad x_t &= 0.4\,(1-\widetilde{\text{recalls}}_t) + 0.3\,(1-\widetilde{\text{warning letters}}_t) + 0.3\,\widetilde{\text{audit pass rate}}_t, \label{eq:xidx}\\
\text{Active regulators:}\quad y_t &= 0.5\,\widetilde{\text{inspection activity}}_t + 0.5\,\widetilde{\text{enforcement activity}}_t, \label{eq:yidx}\\
\text{Public scrutiny:}\quad z_t &= \tfrac{1}{3}\big(\widetilde{\text{litigation}}_t + \widetilde{\text{media coverage}}_t + \widetilde{\text{trust indicators}}_t\big), \label{eq:zidx}\\
\text{Bureaucratic friction:}\quad k_t &= \tfrac{1}{3}\big(\widetilde{\text{approval delay}}_t + \widetilde{\text{compliance cost}}_t + \widetilde{\text{administrative burden}}_t\big). \label{eq:kidx}
\end{align}
These weights encode stated modeling judgments and are held fixed throughout. In \cref{eq:xidx}, recalls carry the largest weight because a recall is an adjudicated, patient-facing failure and is therefore the most direct observable of non-compliance, whereas warning letters and audit pass rates are process indicators that may either precede or follow a realized safety event; the latter two receive equal weight because neither is the stronger signal of realized harm. The complements $(1-\cdot)$ appear because recalls and warning letters increase with undesirable behavior while $x$ measures ethical behavior. In \cref{eq:yidx} detection and prosecution receive equal weight because the inspection-game payoff of \cref{sec:games} depends on both: an inspection never followed by enforcement imposes no expected cost on a defector, and enforcement cannot occur without inspection. \Cref{eq:zidx,eq:kidx} use equal weights because we have no principled basis for ranking their constituent proxies. The series are sampled biennially, giving the eight observations used for calibration (\cref{tab:indices}, Appendix~\ref{app:params}); this smooths single-year reporting noise in the FDA dashboards while retaining the 2020 inflection. Since these weights are chosen rather than measured, the released dataset contains the raw pre-normalization series so that alternative weightings can be substituted.

The sensitivities in Panel~4.4 (\cref{fig:policy}) are likewise computed outside the calibration, by one-factor-at-a-time (OFAT) finite differencing. Due to the dynamics oscillating (\cref{sec:jac}), the state at any single time depends on the phase of the cycle, so we summarize a run by its long-run mean compliance
\begin{equation}
\langle x\rangle \;=\; \frac{1}{300}\int_{200}^{500} x(t)\,\mathrm{d} t ,
\label{eq:meanx}
\end{equation}
integrated from the common initial state $(0.5,0.5,0.5)$ at the historical mean friction $\kbar=0.7975$ (see \cref{sec:provenance}). The window starts at $t=200$, after most of the decay of public scrutiny, and is long enough to contain at least two full cycles (periods are $80$--$160$ time units here). For each parameter $\theta_i$ in turn, holding all others at their fitted values, the plotted bar lengths are the one-sided differences
\begin{equation}
\Delta^{\pm}_i \;=\; \langle x\rangle\!\left(\theta_i (1 \pm 0.2)\right) - \langle x\rangle\!\left(\theta_i\right),
\label{eq:tornado}
\end{equation}
evaluated separately in each direction because the response is asymmetric. The $\pm20\%$ span is a reporting convention rather than an estimated confidence interval, and should be read as a finite local perturbation, not as a probabilistic bound. These sensitivities are diagnostics computed from the fitted model and are not inputs to the calibration.

\subsection{Fixed Points, Stability, and the Friction Thresholds}\label{sec:jac}
\paragraph{Jacobian:} The Jacobian $J_{ij}=\partial\dot x_i/\partial x_j$ of \cref{eq:generic} is
\begin{equation}
J =
\begin{pmatrix}
(1-2x)\,F & x(1-x)\,a_y & x(1-x)\,a_z \\[4pt]
y(1-y)\,b_x & (1-2y)\,G & 0 \\[4pt]
z(1-z)\,c_x & 0 & (1-2z)\,H
\end{pmatrix}.
\label{eq:jacobian}
\end{equation}
The entry $J_{12}=x(1-x)\,a_y = x(1-x)\big[(S_{pol}+\alpha)-k^3\big]$ captures the tension between state incentives and bureaucratic friction.

\paragraph{Fixed points:} A fixed point requires $\dot x=\dot y=\dot z=0$, i.e., each coordinate is $0$ or $1$ or its bracket in \cref{eq:x,eq:y,eq:z} vanishes. Combining the cases gives three families:
\begin{itemize}
\item the eight corners of the cube, $x,y,z\in\{0,1\}$;
\item four face fixed points, one on each of the faces $y=0$, $y=1$, $z=0$, and $z=1$:
\begin{equation}
\begin{aligned}
E_{y0} &= \big(x_y^*,\,0,\,-a_0/a_z\big), & E_y &= \big(x_y^*,\,1,\,-(a_0+a_y)/a_z\big),\\
E_z &= \big(x_z^*,\,-a_0/a_y,\,0\big), & E_{z1} &= \big(x_z^*,\,-(a_0+a_z)/a_y,\,1\big),
\end{aligned}
\label{eq:facefp}
\end{equation}
where the two faces in each pair share an $x$-coordinate,
\begin{equation}
x_z^* \coloneqq -\frac{b_0}{b_x} = 1-\frac{C_r}{B_r},\qquad x_y^* \coloneqq -\frac{c_0}{c_x}=1-\frac{C_p}{B_p}.
\label{eq:xstar}
\end{equation}
A face fixed point is relevant only if its remaining coordinate lies in $[0,1]$. There are no fixed points on the faces $x=0$ and $x=1$ other than corners, because there $G$ and $H$ are nonzero constants;
\item points that exist only for special parameter values. The only non-trivial one is the case $C_r/B_r=C_p/B_p$, i.e., $x_z^*=x_y^*$, in which $G$ and $H$ vanish together and the system has a whole line of interior fixed points, $x=x_z^*$, $z=-(a_0+a_y y)/a_z$.
\end{itemize}
Under the calibrated constants $x_z^*=0.900$ and $x_y^*=0.887$, so the special case does not occur and there is no interior fixed point. \Cref{sec:longrun} shows that every trajectory starting in the interior of the cube therefore approaches its boundary. (One degeneracy of the fit should be noted: the fitted $C_{base}$ and $\gamma$ both sit at the lower bound $0.1$ of the optimizer, so $F=a_0+a_z=0$ along the whole edge $y=0$, $z=1$, which is therefore a line of fixed points containing $E_{y0}$ and $E_{z1}$. Since both coordinates at the edge are at their extremes, this edge plays no role in the dynamics discussed below.)

\paragraph{Corners:} At a corner the off-diagonal entries of \cref{eq:jacobian} vanish, so the eigenvalues are $\pm F$, $\pm G$, $\pm H$ evaluated there (with sign $+$ for a coordinate equal to $0$ and $-$ for a coordinate equal to $1$). They are real and, generically, nonzero, so every corner is a hyperbolic node or saddle. Under the fitted constants and for every $k<k^{**}$ (defined below), all eight corners are saddles in the full system (\cref{tab:fixedpoints}), so there is no source or sink anywhere in the cube. Restricted to the face $x=0$ (the $yz$-plane), the corner $(0,0,0)$ is a source and $(0,1,1)$ a sink, and on the face $x=1$ the corner $(1,1,1)$ is a source and $(1,0,0)$ a sink; this pattern is visible in Panel~2.2 (\cref{fig:basins}).

\paragraph{Face fixed points and the limits of linearization:} At $E_z$ (on $z=0$) the Jacobian has one eigenvalue transverse to the face and a pair inside it,
\begin{equation}
\lambda_\perp = H(x_z^*) = B_p\,(x_y^*-x_z^*),\qquad
\lambda_\parallel^2 = \frac{a_0\,b_0\,(a_0+a_y)(b_0+b_x)}{a_y\,b_x},
\label{eq:eigEz}
\end{equation}
and analogous expressions hold at $E_y$ with $(b_0,b_x,a_y)\to(c_0,c_x,a_z)$ and $a_0\to a_0+a_y$ (Appendix~\ref{app:hamiltonian}). If $\lambda_\parallel^2>0$ the in-face pair is real with opposite signs and the point is a saddle; if $\lambda_\parallel^2<0$ the pair is purely imaginary, $\lambda_\parallel=\pm i\omega$, and the linearization predicts a center within the face. For $E_z$ with $0<y^*<1$, i.e., $a_0<0<a_0+a_y$ and $b_0>0>b_0+b_x$, the expression is negative, so $E_z$ is a center whenever it lies in the domain. 

Centers are not hyperbolic fixed points (Appendix~\ref{app:hamiltonian} gives the Jacobian and eigenvalues at every fixed point), so the Hartman--Grobman theorem does not apply: higher-order terms dropped by the linearization could in principle turn a linear center into a weakly stable or unstable spiral \cite{strogatz2024}. A separate argument is needed. On the face $z=0$ the system has the conserved quantity
\begin{equation}
\mathcal{H}(x,y) \;=\; \frac{x^{\,b_0}\,(1-y)^{\,a_0+a_y}}{(1-x)^{\,b_0+b_x}\,y^{\,a_0}},\qquad \frac{\mathrm{d}\mathcal{H}}{\mathrm{d}t}=0 ,
\label{eq:H}
\end{equation}
derived in Appendix~\ref{app:hamiltonian}. Under the conditions above, $\ln\mathcal{H}\to-\infty$ on all four edges of the face and has a single maximum at $E_z$, so every level set is a closed curve around $E_z$: every orbit on the face is periodic, and $E_z$ is a true (nonlinear) center \cite[Sec.~6.5]{strogatz2024}. The same holds for $E_y$ on the face $y=1$. With the calibrated constants at the 2010 friction $k=0.60$, $E_z=(0.900,\,0.320,\,0)$ with $\lambda_\parallel=\pm0.078i$ and $\lambda_\perp=-0.0119$; small orbits have period $2\pi/\omega\approx80$ time units and the orbit of \cref{fig:core} has period $\approx105$. Since one time unit corresponds to roughly one calendar year in the calibration (\cref{sec:ml}), a single firm--regulator cycle spans decades.

\paragraph{The friction thresholds:} Only $a_y=S_{pol}+\alpha-k^3$ depends on friction, and two values of $k$ organize the dynamics. Tracking $E_z$, its $y$-coordinate $y^*=C_{base}/(S_{pol}+\alpha-k^3)$ increases with $k$ and reaches the face $y=1$ when
\begin{equation}
k = k^* \coloneqq (S_{pol}+\alpha-C_{base})^{1/3} \approx 0.754 .
\label{eq:kstar}
\end{equation}
At this point $a_0+a_y$ changes sign, $\lambda_\parallel^2$ in \cref{eq:eigEz} becomes positive, and the center becomes a saddle outside the cube. Below this threshold the condition $S_{pol}+\alpha>k^3$ holds automatically: $y^*<1$ requires $a_y=S_{pol}+\alpha-k^3>C_{base}>0$, and then the in-face Jacobian at $E_z$ has zero trace and determinant $x^*(1-x^*)\,y^*(1-y^*)\,B_r\,a_y>0$, the linear signature of a center. The condition is therefore not a separate assumption of the model but a consequence of $E_z$ lying in the domain. (At $k=(S_{pol}+\alpha)^{1/3}$ the coordinate $y^*$ passes through infinity and $E_z$ turns back into a center, now at $y^*<0$, which is outside the domain and has no effect on the dynamics.) As $E_z$ leaves through $y=1$, the face fixed point $E_y=(x_y^*,\,1,\,z_y^*)$ with $z_y^*=(C_{base}-a_y)/\gamma$ enters the face $y=1$ through its edge $z=0$; it is a center within that face and remains in the domain until $z_y^*$ reaches $1$, at
\begin{equation}
k = k^{**} \coloneqq (S_{pol}+\alpha+\gamma-C_{base})^{1/3} \approx 0.809 ,
\label{eq:k2star}
\end{equation}
which under the fitted equality $\gamma=C_{base}$ coincides with $(S_{pol}+\alpha)^{1/3}$. For $k>k^{**}$ and the fitted constants, $F=-C_{base}+a_y y+\gamma z\le \gamma-C_{base}=0$ throughout the cube, with equality only on the degenerate edge, so $\dot x<0$ in the interior and compliance decays everywhere. The corner $(0,1,1)$, with no ethical firms and regulators and the public fully active, then becomes the only sink (eigenvalues $a_y$, $-b_0$, $-c_0$, all negative). No Hopf bifurcation occurs at either threshold: the centers are neutral for every $k$ at which they exist, and they leave the domain through a face of the cube rather than by changing stability.

\begin{table}[H]
\centering
\caption{Fixed points under the calibrated constants (\cref{tab:params}) at the 2010 friction $k=0.60$ and the 2020 friction $k=0.88$. Eigenvalues are of the full $3\times3$ Jacobian \eqref{eq:jacobian}. The corners $(0,0,1)$ and $(1,0,1)$ and the face points $E_{y0}$, $E_{z1}$ lie on the degenerate edge $y=0$, $z=1$ (one zero eigenvalue, because the fitted $\gamma=C_{base}$) and are omitted.}\label{tab:fixedpoints}
{\small
\begin{tabular}{@{}lccc@{}}
\toprule
Point & Eigenvalues at $k=0.60$ & Type at $k=0.60$ & Type at $k=0.88$ \\
\midrule
$(0,0,0)$ & $-0.100,\ 0.900,\ 0.796$ & saddle & saddle \\
$(0,1,0)$ & $0.213,\ -0.900,\ 0.796$ & saddle & saddle \\
$(0,1,1)$ & $0.313,\ -0.900,\ -0.796$ & saddle & \textbf{sink} \\
$(1,0,0)$ & $0.100,\ -0.100,\ -0.102$ & saddle & saddle \\
$(1,1,0)$ & $-0.213,\ 0.100,\ -0.102$ & saddle & saddle \\
$(1,1,1)$ & $-0.313,\ 0.100,\ 0.102$ & saddle & source \\
$E_z=(0.900,\,0.320,\,0)$ & $\pm0.078i,\ -0.012$ & center in face $z=0$ & outside cube \\
$E_y=(0.887,\,1,\,z_y^*)$ & --- & outside cube ($z_y^*<0$) & outside cube ($z_y^*>1$) \\
\bottomrule
\end{tabular}}
\end{table}

\subsection{Long-Run Behavior: Approach to the Boundary and Absence of Chaos}\label{sec:longrun}
The arguments in this subsection are expository rather than formal proofs; they explain the behavior that the simulations of \cref{sec:validation} display.

\paragraph{Every interior trajectory approaches the boundary:} Because $\tfrac{\mathrm{d}}{\mathrm{d}t}\ln[y/(1-y)]=G$ and $\tfrac{\mathrm{d}}{\mathrm{d}t}\ln[z/(1-z)]=H$, the quantity
\begin{equation}
V \;=\; \frac{1}{B_r}\ln\frac{y}{1-y} \;-\; \frac{1}{B_p}\ln\frac{z}{1-z}
\qquad\text{obeys}\qquad
\dot V \;=\; \frac{G}{B_r}-\frac{H}{B_p} \;=\; \frac{C_p}{B_p}-\frac{C_r}{B_r} \;=\; x_z^*-x_y^*
\label{eq:V}
\end{equation}
exactly, for every state. The $x$-dependence cancels because regulators and the public respond to the same signal, the share $1-x$ of defecting firms. Under the calibrated constants $\dot V=0.0133>0$, so $V$ grows linearly without bound and every interior trajectory must approach the face $y=1$ or the face $z=0$ (or both). This is the precise sense in which the absence of an interior fixed point forces one population to a boundary: the regulators, whose cost-to-benefit ratio $C_r/B_r=0.100$ is lower, out-compete the public ($C_p/B_p=0.113$) as enforcers. If instead $C_r/B_r=C_p/B_p$, $V$ is conserved, which is the line of fixed points noted in \cref{sec:jac}.

\paragraph{Which boundary:} For $k<k^*$ the face $y=1$ cannot hold trajectories: there $F=a_0+a_y+a_z z>0$, so compliance rises until $x>x_z^*$, after which $G<0$ pushes $y$ away from $1$. The trajectories therefore approach the face $z=0$, and on that face they follow the closed orbits of \cref{eq:H}, so the long-run motion is a neutral firm--regulator cycle around $E_z$ with public scrutiny dying out. While the orbit stays near the face, $\ln[y/(1-y)]$ returns to roughly the same value each cycle, so by \cref{eq:V} the log-odds of $z$ fall by approximately $B_p(x_z^*-x_y^*)\,T\approx0.0119\,T$ per cycle of period $T$: $z$ shrinks by a factor of about $e^{-0.0119\times105}\approx0.29$ per cycle (\cref{fig:zdecay}). A simpler but more restrictive sufficient condition is also available: since $z(1-z)>0$, the sign of $\dot z$ is the sign of $c_0+c_x x$, so any trajectory that stays to the right of the plane $x=x_y^*$ for all time has $\dot z<0$ throughout and decays monotonically to $z=0$. For $k^*<k<k^{**}$ the face $z=0$ cannot hold trajectories, since there $F=a_0+a_y y<0$ drives $x$ toward $0$, where $H=c_0>0$ makes $z$ grow; trajectories instead approach the face $y=1$ and circulate around $E_y$, so regulators saturate and public scrutiny carries enforcement. For $k>k^{**}$ trajectories converge to the corner $(0,1,1)$.

\begin{figure}[H]
    \centering
    \includegraphics[width=\textwidth]{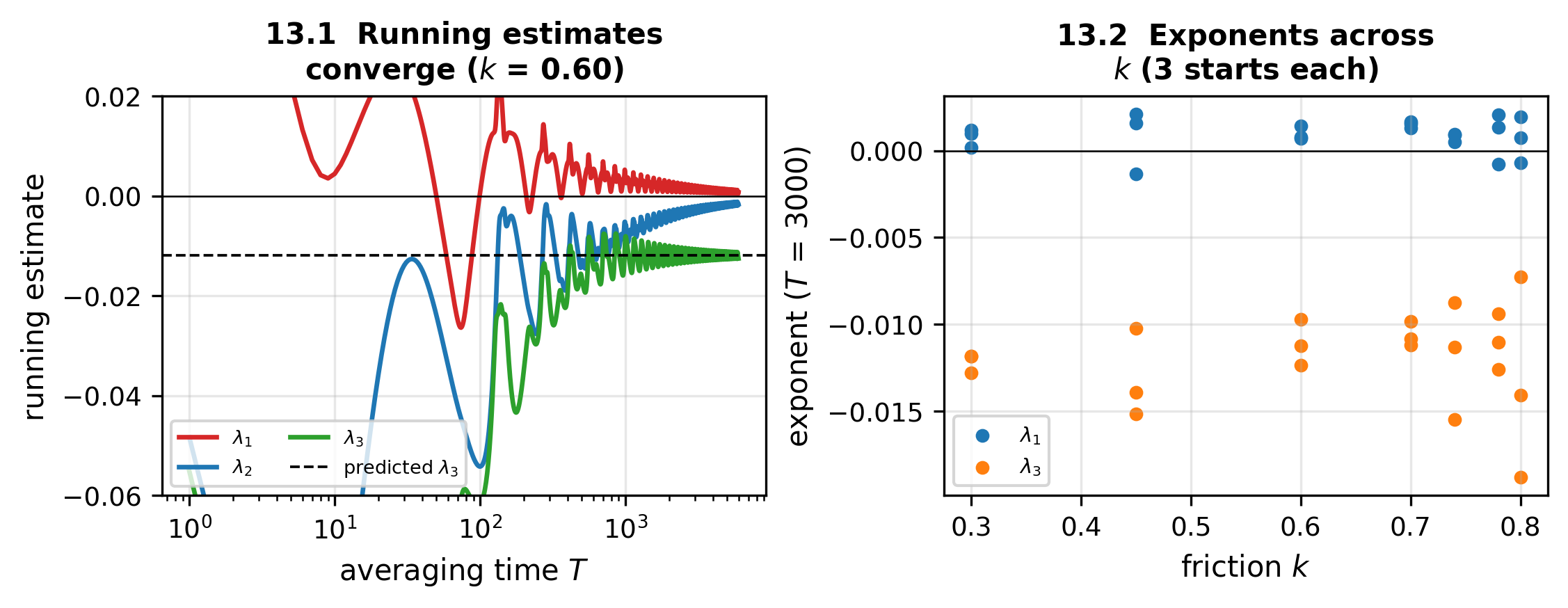}
    \caption{\textbf{Lyapunov exponents} (Benettin QR method, re-orthonormalized every time unit). (13.1) running estimates of $\lambda_1\ge\lambda_2\ge\lambda_3$ from $(0.5,0.5,0.5)$ at $k=0.60$ against the averaging time $T$, with the predicted transverse rate $B_p(x_y^*-x_z^*)$ (dashed); (13.2) $\lambda_1$ and $\lambda_3$ at $T=3000$ for three random initial conditions (uniform on $[0.05,0.95]^3$, seed 1) at each of seven values of $k$.}\label{fig:lyapunov}
\end{figure}

\paragraph{Absence of chaos:} A two-dimensional autonomous flow cannot be chaotic (Poincar\'e--Bendixson theorem), but a three-dimensional one can \cite{strogatz2024}. We therefore computed the three Lyapunov exponents numerically with the Benettin QR method; a positive exponent indicates chaos \cite{strogatz2024}. As an internal accuracy check, any bounded trajectory that does not terminate at a fixed point has one exponent identically zero (the direction along the flow) \cite{haken1983}. From initial condition $(0.5,0.5,0.5)$ at $k=0.60$, averaging over $T=10$ (about one tenth of an orbital period) gives Lyapunov exponents $(0.0044,\,-0.1016,\,-0.1427)$; these short-time values are dominated by the transient. Averaging over $T=6000$ gives $(0.0007,\,-0.0017,\,-0.0126)$: two exponents converge to zero (the flow direction and the neutral direction across the nested orbits) and the third to the transverse rate $B_p(x_y^*-x_z^*)=-0.0119$ predicted by \cref{eq:V} (\cref{fig:lyapunov}). A coarse sweep over $k\in[0.3,0.8]$ with three random initial conditions per value gave a largest exponent of at most $0.002$. We did not perform a full sweep over parameter space, so we cannot exclude chaos everywhere, but we find no evidence of it.

\paragraph{Hamiltonian structure:} The face dynamics are not Hamiltonian in the coordinates $(x,y)$: the divergence $\partial_x\dot x+\partial_y\dot y$ is nonzero, so phase-space area is not preserved, even though $\mathcal{H}$ is conserved. The system is instead \emph{conformally} Hamiltonian \cite{garcia2020}, becoming a canonical Hamiltonian system after a state-dependent rescaling of time; equivalently, in log-odds coordinates $u=\ln[x/(1-x)]$, $v=\ln[y/(1-y)]$ it is exactly Hamiltonian with Hamiltonian $-\ln\mathcal{H}$, a known property of replicator dynamics for $2\times2$ bimatrix games \cite{hofbauer1996}. Appendix~\ref{app:hamiltonian} provides the details.

\subsection{Parameter Origin}\label{sec:provenance}
Every numerical constant in this study belongs to one of three declared classes, and all values are listed in Appendix~\ref{app:params}. \textbf{(i) Fitted:} the eight payoff constants ($C_{base}$, $S_{pol}$, $\alpha$, $\gamma$, $B_r$, $C_r$, $B_p$, $C_p$) are estimated by the bounded L-BFGS-B calibration of \cref{sec:ml}; four of them ($C_{base}$, $C_r$, $\gamma$ at their lower bounds, $B_r$ at its upper bound) finish on a bound (\cref{tab:params}). \textbf{(ii) Empirical:} the friction series $k(t)$, the initial condition $(x_0,y_0,z_0)=(0.65,0.20,0.15)$ (the 2010 row of the composite indices), and the index-construction weights of \cref{sec:params} are computed from the source datasets of \cref{sec:data}. Two reference friction values recur: $k=0.60$, the 2010 value, which lies below $k^*$ and is used for the core dynamics (\cref{fig:core,fig:basins,fig:stoch}) and the phase-plane analysis; and the 2010--2024 mean $\kbar=0.7975$, which lies between $k^*$ and $k^{**}$ (and $1.4\%$ below $k^{**}$) and is used for the policy experiments (\cref{fig:policy,fig:heatmaps}), so that the levers studied can push the system either way. Results computed at $\kbar$ are therefore threshold sensitivities by construction. \textbf{(iii) Illustrative:} constants used only in robustness demonstrations are not inputs to any headline result: the noise amplitude $\sigma=0.08$ and the log-normal shock scale $0.5$ (order-of-magnitude illustrations \cite{oksendal2003}), the spatial temptation payoff $b=1.65$ (inside the range $1<b<2$ studied by Nowak and May \cite{nowak1992}), the $100\times100$ lattice with 10\% initial defectors, the $\pm20\%$ perturbation convention of the sensitivity analysis, and the regulator audit delay $\tau_y=1.5$ years, which approximates the interval between FDA surveillance inspections of a given establishment; for comparison, until the FDA Safety and Innovation Act of 2012 introduced a risk-based schedule, the Federal Food, Drug, and Cosmetic Act required every registered drug establishment to be inspected at least once in every two-year period \cite{fdca510h}.

\subsection{Numerical Methods: Calibration and Model Extensions}\label{sec:ml}
To fit the ODEs to the empirical indices we used the \texttt{L-BFGS-B} (limited-memory Broyden--Fletcher--Goldfarb--Shanno, bound-constrained) quasi-Newton algorithm \cite{byrd1995}, chosen because it enforces simple bounds on each parameter. During calibration the friction enters as a forcing: $k(t)$ is linearly interpolated between the biennial observations, with one time unit equal to one calendar year. The objective is a weighted mean squared error over the eight observation years, in which the compliance residuals are weighted eight times more heavily than the regulator and public residuals and receive a linearly increasing time weight from $1.0$ (2010) to $3.0$ (2024), to prioritize the recent compliance decline. All other simulations hold $k$ constant and integrate \cref{eq:generic} in log-odds coordinates, in which the right-hand side is bounded and shares that become exponentially small near a face are represented accurately.

The spatial extension is a deterministic cellular automaton on a $100\times100$ lattice with periodic (wrap-around) boundary conditions, a standard setting for evolutionary games on graphs \cite{szabo2007}. Each cell is ethical or defecting; in each generation an ethical cell earns $1$ per ethical neighbor and a defecting cell earns $b$ per ethical neighbor in its Moore neighborhood (the eight surrounding cells), following the weak prisoner's dilemma of Nowak and May \cite{nowak1992}. Every cell then synchronously adopts the strategy of the highest-earning cell among itself and its neighbors (an imitate-the-best update rule). This extension is illustrative and does not use the fitted constants.

For the stochastic extension we add multiplicative noise to the firm equation, $\mathrm{d} x = \dot x\,\mathrm{d} t + \sigma\,x(1-x)\,\mathrm{d} W_t$, where $W_t$ is a standard Wiener process \cite{oksendal2003}, and integrate with the Euler--Maruyama scheme (step $\Delta t=0.15$, noise increments scaled by $\sqrt{\Delta t}$). The factor $x(1-x)$ is the geometric-Brownian-motion form $\sigma x\,\mathrm{d} W$ adapted so that the faces $x=0,1$ remain invariant.

Finally, we model regulatory lag with a delay differential equation (DDE) \cite{kuang1993}, integrated with an explicit Euler scheme. Regulators respond to the firm state delayed by $\tau_y = 1.5$ years, while firms and the public respond to the current state:
\begin{align}
\dot{x}(t) &= x(1-x)\big[-(C_{base}+k^3 y) + y(S_{pol}+\alpha) + z\gamma\big], \label{eq:ddex}\\
\dot{y}(t) &= y(1-y)\big[\big(1-\boldsymbol{x(t-\tau_y)}\big)\,B_r - C_r\big], \label{eq:ddey}\\
\dot{z}(t) &= z(1-z)\big[(1-x)\,B_p - C_p\big]. \label{eq:ddez}
\end{align}
The only change from \crefrange{eq:x}{eq:z} is the delayed argument $x(t-\tau_y)$ in the regulator equation (bold in \cref{eq:ddey}). The code for all numerical methods and figures is available in the project repository (see Data and Code Availability).
\section{Results: Model Dynamics and Visualizations}\label{sec:results}
This section presents the simulations. Unless a caption states otherwise, every run uses the calibrated constants of \cref{tab:params}, constant friction $k$, and evolutionary time $t$ in units of approximately one calendar year. Panel numbers ($n.m$) refer to the labels printed on each panel.

\begin{figure}[H]
    \centering
    \includegraphics[width=\textwidth]{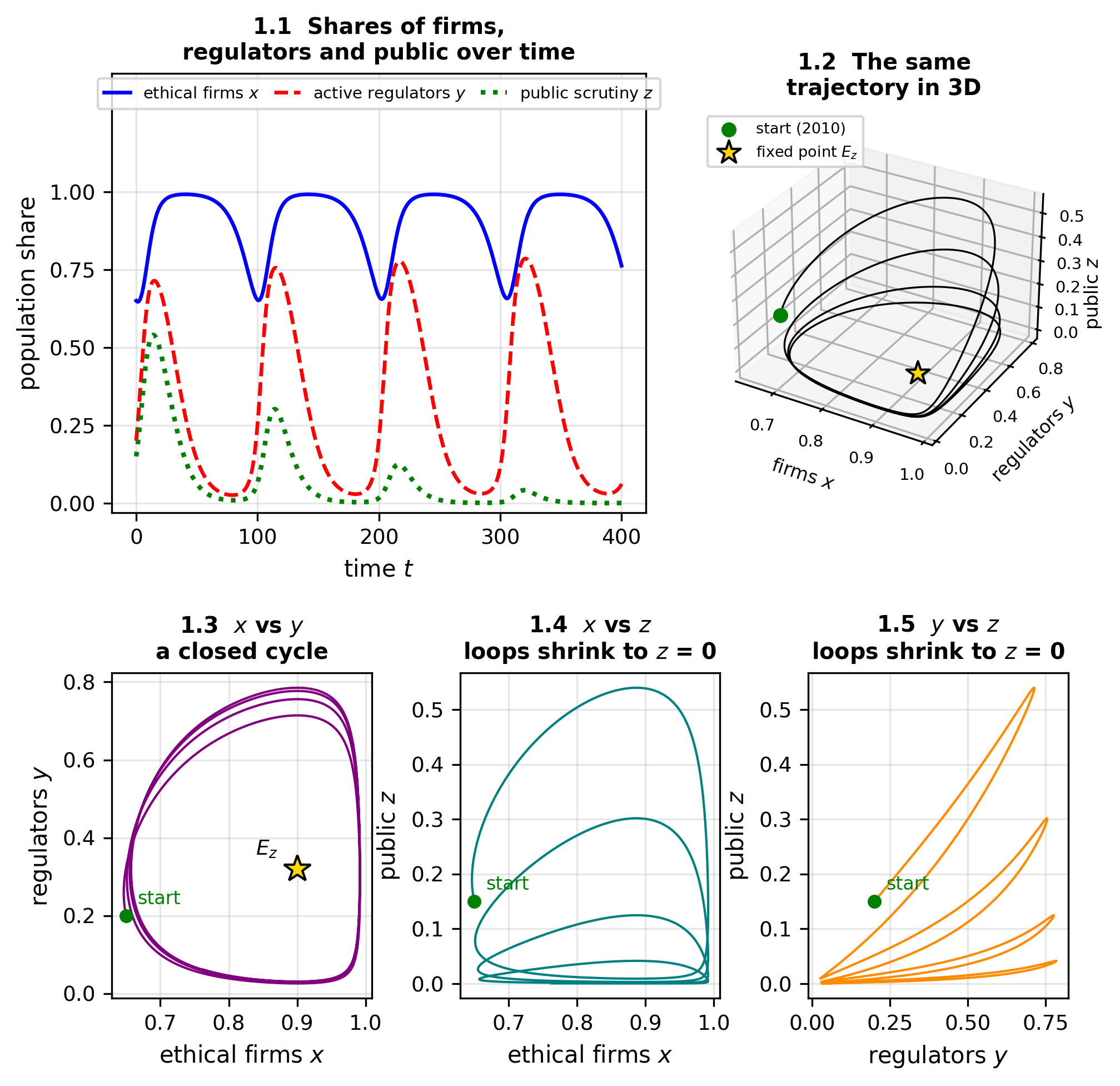}
    \caption{\textbf{Core dynamics:} One trajectory of \crefrange{eq:x}{eq:z} from the 2010 state $(x_0,y_0,z_0)=(0.65,0.20,0.15)$ at the 2010 friction $k=0.60$, for $0\le t\le400$. (1.1) shares of ethical firms $x$, active regulators $y$, and public scrutiny $z$ over time; (1.2) the same trajectory as a 3D orbit; (1.3--1.5) its projections onto the firm--regulator, firm--public, and regulator--public planes, each scaled to the range its two variables cover. The green dot marks the starting state and the gold star the face fixed point $E_z=(0.900,0.320,0)$ at its analytic coordinates (\cref{eq:facefp}); $E_z$ is drawn only where it is meaningful, in the 3D view and in the firm--regulator plane, since it is the center of the loops in 1.3.}\label{fig:core}
\end{figure}
\textbf{Analysis of \cref{fig:core} (core dynamics):} \textbf{Panel 1.1} shows a persistent cycle: compliance oscillates between about $0.65$ and $0.99$, and regulator activity between $0.03$ and $0.78$, with successive regulator peaks $100$--$105$ time units apart. The mechanism is the predator--prey structure of the inspection game. When compliance exceeds $x_z^*=1-C_r/B_r=0.90$, inspection no longer pays ($\dot y<0$) and regulators withdraw; with fewer inspections defection becomes profitable and compliance falls; once compliance drops below $0.90$, inspection pays again and regulators return, restoring compliance. Public scrutiny follows the same rhythm but its peaks shrink from cycle to cycle ($0.54$, $0.30$, $0.12$, $0.04$), as predicted by \cref{eq:V}. \textbf{Panel 1.2} shows the orbit descending onto the face $z=0$ while circling $E_z$, and \textbf{Panel 1.3} shows that in the firm--regulator plane the loops converge onto a single closed orbit around $E_z$ (a neutral cycle, \cref{sec:jac}). \textbf{Panels 1.4 and 1.5} show the same convergence in the planes that include $z$: each loop sits lower than the previous one. The significance of this figure is that, in the model, ethical compliance is a recurring cycle rather than a permanently achieved state: near-universal compliance removes the incentive to inspect, and so it is not sustained.

\begin{figure}[H]
    \centering
    \includegraphics[width=\textwidth]{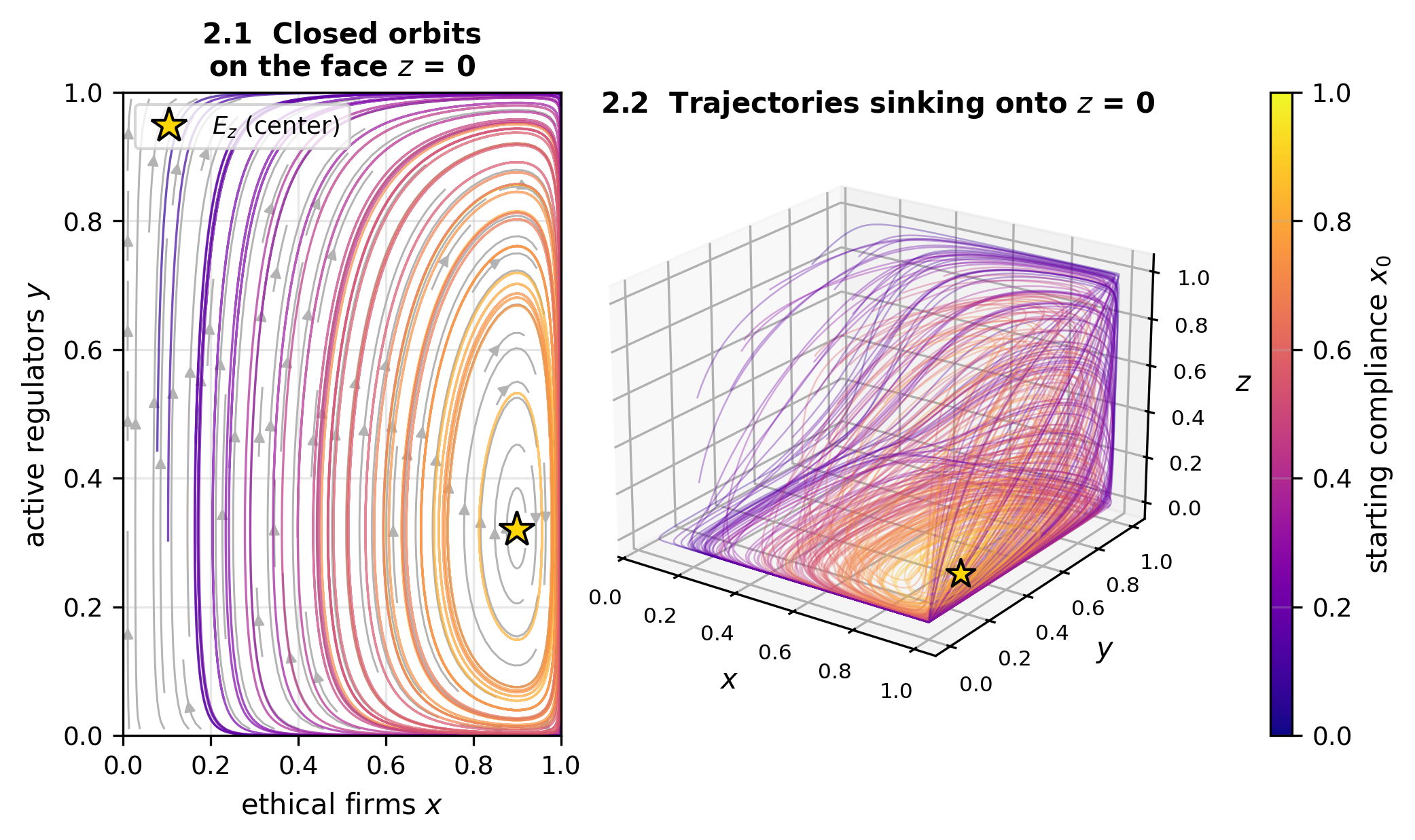}
    \caption{\textbf{Phase portraits} at $k=0.60$. (2.1) the velocity field $(\dot x,\dot y)$ on the invariant face $z=0$ (gray streamlines) with $40$ trajectories from random initial conditions on that face (uniform on $[0.05,0.95]^2$, seed 42), integrated to $t=300$; (2.2) $60$ trajectories from random initial conditions in the cube (uniform on $[0.05,0.95]^3$), integrated to $t=600$. In both panels a trajectory's color shows its starting compliance $x_0$ (color bar). Gold star: $E_z$.}\label{fig:basins}
\end{figure}
\textbf{Analysis of \cref{fig:basins} (phase portraits):} \textbf{Panel 2.1} shows that every trajectory on the face $z=0$ is a closed orbit around $E_z$; different starting points lie on different orbits (the level sets of $\mathcal{H}$, \cref{eq:H}), and none spirals in or out. Even an industry that starts with almost no ethical firms ($x\approx0$, the left side of the panel) returns to high compliance: with few ethical firms, inspection pays, regulators multiply, and once $y$ exceeds $y^*=0.32$ the fitness advantage of compliance turns positive. \textbf{Panel 2.2} shows the corresponding behavior in the full cube. Trajectories starting at all three public-scrutiny levels descend onto the face $z=0$ and wrap around $E_z$; those that start with few ethical firms first move toward the corner $(0,1,1)$, the sink of the face $x=0$, before turning toward high compliance. The panel thus shows the two features established in \cref{sec:jac,sec:longrun}: neutral cycling within the face $z=0$ and slow decay of $z$ toward it.

\begin{figure}[H]
    \centering
    \includegraphics[width=0.86\textwidth]{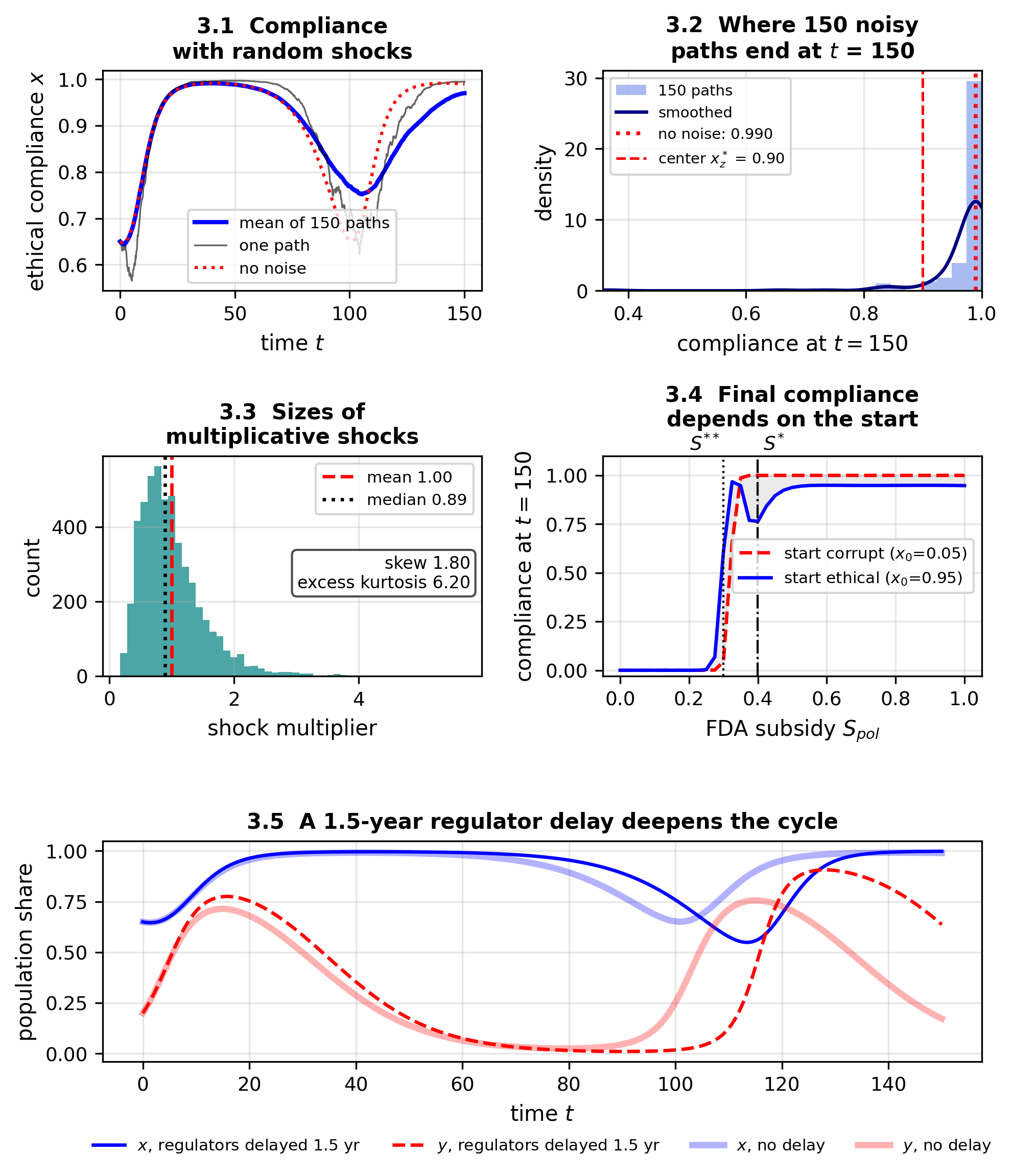}
    \caption{\textbf{Noise, path dependence, and delay:} (3.1) $150$ Euler--Maruyama paths of the firm equation with multiplicative noise $\sigma x(1-x)\,\mathrm{d} W$, $\sigma=0.08$, $\Delta t=0.15$ (seed 7), from $(0.65,0.20,0.15)$ at $k=0.60$, $0\le t\le150$: mean path, one sample path, and the deterministic path; (3.2) histogram and kernel density estimate of $x(150)$ across the $150$ paths, with the deterministic $x(150)$ and the face fixed point $x_z^*$; (3.3) $5000$ draws of a log-normal shock multiplier with unit mean and log-scale $0.5$; (3.4) $x(150)$ against the subsidy $S_{pol}$ from a corrupt start $(0.05,0.2,0.2)$ and an ethical start $(0.95,0.2,0.2)$ at $k=\kbar=0.7975$, with $S^{**}=\kbar^3+C_{base}-\alpha-\gamma=0.30$ (the subsidy at which $\kbar=k^{**}$) and $S^*=\kbar^3+C_{base}-\alpha=0.40$ (at which $\kbar=k^*$) marked; (3.5) the delayed system \crefrange{eq:ddex}{eq:ddez} with $\tau_y=1.5$ (thin lines) against the undelayed system (thick, faded lines), from $(0.65,0.20,0.15)$ at $k=0.60$.}\label{fig:stoch}
\end{figure}
\textbf{Analysis of \cref{fig:stoch} (noise, path dependence, and delay):} Deterministic ODEs neglect random shocks, so this figure adds stochastic, historical, and delayed effects. \textbf{Panel 3.1} applies the multiplicative-noise scheme of \cref{sec:ml}, representing clinical-trial failures and market shocks. The noisy paths follow the same cycle as the deterministic one; noise mainly shifts their phase, so the mean path is a damped copy of the deterministic cycle (its minimum is about $0.75$ rather than $0.65$). \textbf{Panel 3.2} shows the distribution of compliance at $t=150$. Most paths end near the deterministic value $x(150)=0.990$ (median $0.992$), with a long left tail down to $0.37$ from paths whose phase was shifted by the noise. All values lie in $[0,1]$. The dashed line at $x_z^*=0.90$ marks the center of the cycle, which is also the time-average of $x$ over any closed orbit (\cref{sec:validation}), not a point where paths accumulate. \textbf{Panel 3.3} shows the shape assumed for multiplicative financial shocks: the distribution is strongly right-skewed, with the mean ($1.00$) above the median ($0.89$), skewness $1.80$, and excess kurtosis $6.20$, so a few large losses dominate. These statistics follow from the log-normal assumption by construction; the panel illustrates the assumption rather than providing evidence for it.

\textbf{Panel 3.4} examines path dependence. Path dependence (hysteresis in the broad sense) means that the state a system reaches depends on its history, not only on the current value of a parameter. Both starting states collapse when the subsidy is below $S^{**}=0.30$, the value at which $\kbar$ equals $k^{**}$. Between $S^{**}$ and $S^*=0.40$ the outcome depends on the start: compliance at $t=150$ first exceeds $0.9$ at $S_{pol}=0.325$ from the ethical start but only at $S_{pol}=0.35$ from the corrupt start. Above $S^*$ both end high, but at different points on different closed orbits ($x(150)=0.95$ versus $1.00$). In this model path dependence has two sources. First, the conserved quantity $\mathcal{H}$ means that each starting point stays on its own orbit forever, so the memory of the initial condition never fades. Second, because the face $x=0$ is invariant, a market close to collapse recovers only slowly after the subsidy is raised: the log-odds of $x$ grow at the rate $F$, so the recovery time scales like $\ln(1/x_0)/F$. This is not bistability between two attractors, since the model has no attractors below $k^{**}$. It does support a weaker policy point: within a given horizon, re-establishing compliance in a corrupted market requires a somewhat larger subsidy than maintaining it in an ethical one. Finally, \textbf{Panel 3.5} applies the delayed-response system of \crefrange{eq:ddex}{eq:ddez}, in which regulators react to the firm state of $1.5$ years earlier. The delay makes regulators respond late and overshoot (their second peak is $0.91$ instead of $0.76$) and deepens the following compliance trough ($0.55$ instead of $0.65$); the two compliance paths differ by up to $0.34$ within $t\le150$. A lagged regulator therefore over-corrects and amplifies the cycle.

\begin{figure}[H]
    \centering
    \includegraphics[width=0.82\textwidth]{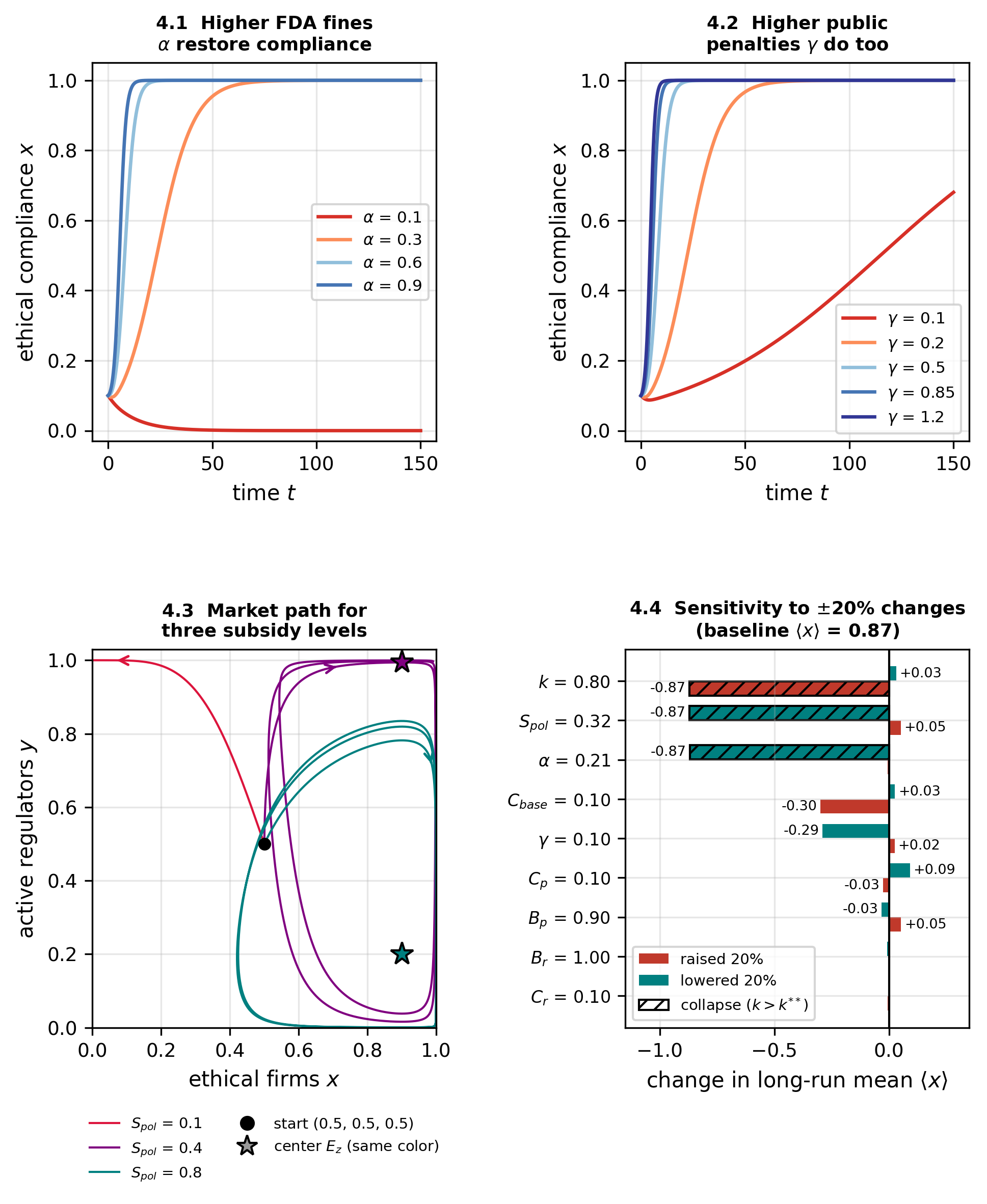}
    \caption{\textbf{Policy experiments and sensitivity} at the historical mean friction $k=\kbar=0.7975$, with all constants at their fitted values unless varied. (4.1) compliance over $0\le t\le150$ from a mostly non-compliant start $(0.1,0.2,0.2)$ for FDA fines $\alpha\in\{0.1,0.3,0.6,0.9\}$ (fitted $\alpha=0.21$); (4.2) the same for public penalties $\gamma\in\{0.1,0.2,0.5,0.85,1.2\}$ (fitted $\gamma=0.1$); (4.3) three trajectories over $0\le t\le400$, all starting at $(x_0,y_0,z_0)=(0.5,0.5,0.5)$ (black dot) and differing only in the subsidy $S_{pol}\in\{0.1,0.4,0.8\}$, drawn in the firm--regulator plane; arrows give the direction of travel and stars the center $E_z$ for each subsidy (none exists for $S_{pol}=0.1$); (4.4) change in the long-run mean compliance $\langle x\rangle$ (\cref{eq:meanx}) when each constant is raised (red) or lowered (teal) by $20\%$ with the others fixed, from $(0.5,0.5,0.5)$ (baseline $\langle x\rangle=0.87$); hatched bars are runs that collapse because they push the system past $k^{**}$, and changes smaller than $0.01$ are not labeled. The baseline of $0.87$ is close to $x_y^*=0.887$, the time average of $x$ on the face $y=1$, where the fitted market cycles at $\kbar$ (see text).}\label{fig:policy}
\end{figure}
\textbf{Analysis of \cref{fig:policy} (policy experiments):} These experiments are run at $\kbar$, which lies between $k^*$ and $k^{**}$ (\cref{sec:provenance}), so the fitted constants put the market in the regime where regulators are fully active and the levers below can tip it either way. \textbf{Panel 4.1} starts from a market with only $10\%$ ethical firms. With a fine of $\alpha=0.1$, $S_{pol}+\alpha+\gamma-C_{base}<\kbar^3$, so $\kbar$ exceeds $k^{**}$ and compliance falls to zero: defection is cheaper than complying even after the fine. A fine of $0.3$ or more pushes compliance to $1.0$ by $t=150$; larger fines get there faster ($\alpha=0.9$ within about $10$ time units, $\alpha=0.3$ within about $50$). \textbf{Panel 4.2} shows that the public penalty works in the same direction as the fine: every value shown lets compliance recover, and larger penalties make the recovery faster. At the fitted $\gamma=0.1$ compliance has only reached $0.68$ by $t=150$, because at $\kbar$ the public is the population that carries enforcement (\cref{sec:longrun}) and a small penalty gives it little leverage. The two levers are not interchangeable in size: $\gamma$ multiplies the public share $z$ and $\alpha$ the regulator share $y$ in \cref{eq:x}, so which one matters more depends on which population is active.

\textbf{Panel 4.3} shows what the subsidy does to the path of the market. Each curve is a single trajectory over $0\le t\le400$, started from $x_0=y_0=z_0=0.5$ (public scrutiny starts at $0.5$, not at $0$) and drawn in the firm--regulator plane. With $S_{pol}=0.1$, below $S^{**}$, the market collapses to the corner $(0,1,1)$: compliance goes to zero while regulators and the public mobilize fully. With $S_{pol}=0.4$, just above $S^*$, the center $E_z$ sits at $y^*=0.99$ and the market cycles close to the face $y=1$, with compliance between $0.50$ and $1.00$. With $S_{pol}=0.8$, $E_z$ moves down to $y^*=0.20$ and the market cycles around it, with compliance between $0.42$ and $1.00$. Raising the subsidy therefore turns collapse into a cycle and moves the center of that cycle to lower regulatory effort ($y^*=C_{base}/(S_{pol}+\alpha-k^3)$), but it does not make the swings in compliance smaller: in a neutral cycle the size of the loop is set by where the market starts (\cref{sec:jac}), not by the subsidy.

\textbf{Panel 4.4} reports the one-at-a-time changes of \cref{eq:tornado}, summarized by the long-run mean compliance $\langle x\rangle$, which averages over the cycle and so does not depend on where in the cycle a run happens to end. The baseline, $\langle x\rangle=0.87$, is not a tuning choice; it follows from the model. At $\kbar$ the fitted market ends up cycling on the face $y=1$ (regulators fully active) around the center $E_y$, and on that face the same time-average argument used in \cref{sec:validation} gives a mean compliance over whole cycles of $x_y^*=1-C_p/B_p=0.887$. The computed $0.871$ is slightly lower because the averaging window ($200\le t\le500$) starts while the trajectory from $(0.5,0.5,0.5)$ is still approaching that face. Three changes collapse compliance ($\Delta\approx-0.87$, hatched): raising friction by $20\%$ (to $k=0.96$), or cutting the subsidy or the fine by $20\%$. Each of these pushes the system past $k^{**}$. The opposite, favorable changes raise $\langle x\rangle$ by only $0.03$--$0.05$. This is the same time-average identity at work: as long as a change leaves the market cycling on the same face, the long-run mean stays close to $x_y^*\approx0.9$, so it has little room to rise. The one larger gain ($+0.09$, from lowering $C_p$) is the exception that fits the mechanism. Lowering $C_p$ by $20\%$ makes the public's cost-to-benefit ratio smaller than the regulators', which reverses the sign of $\dot V$ in \cref{eq:V} and moves the market to a different face where that ceiling no longer applies. Raising $C_{base}$ or lowering $\gamma$ by $20\%$, which breaks the fitted equality $\gamma=C_{base}$, lowers $\langle x\rangle$ by about $0.3$; every other change moves it by less than $0.1$. The asymmetry, with compliance far more sensitive to adverse than to favorable changes, follows from the baseline sitting $1.4\%$ below $k^{**}$: adverse changes cross the threshold, favorable ones only move the system within the cycling regime. Near the threshold, then, heavy red tape collapses compliance faster than subsidies can restore it.

\begin{figure}[H]
    \centering
    \includegraphics[width=\textwidth]{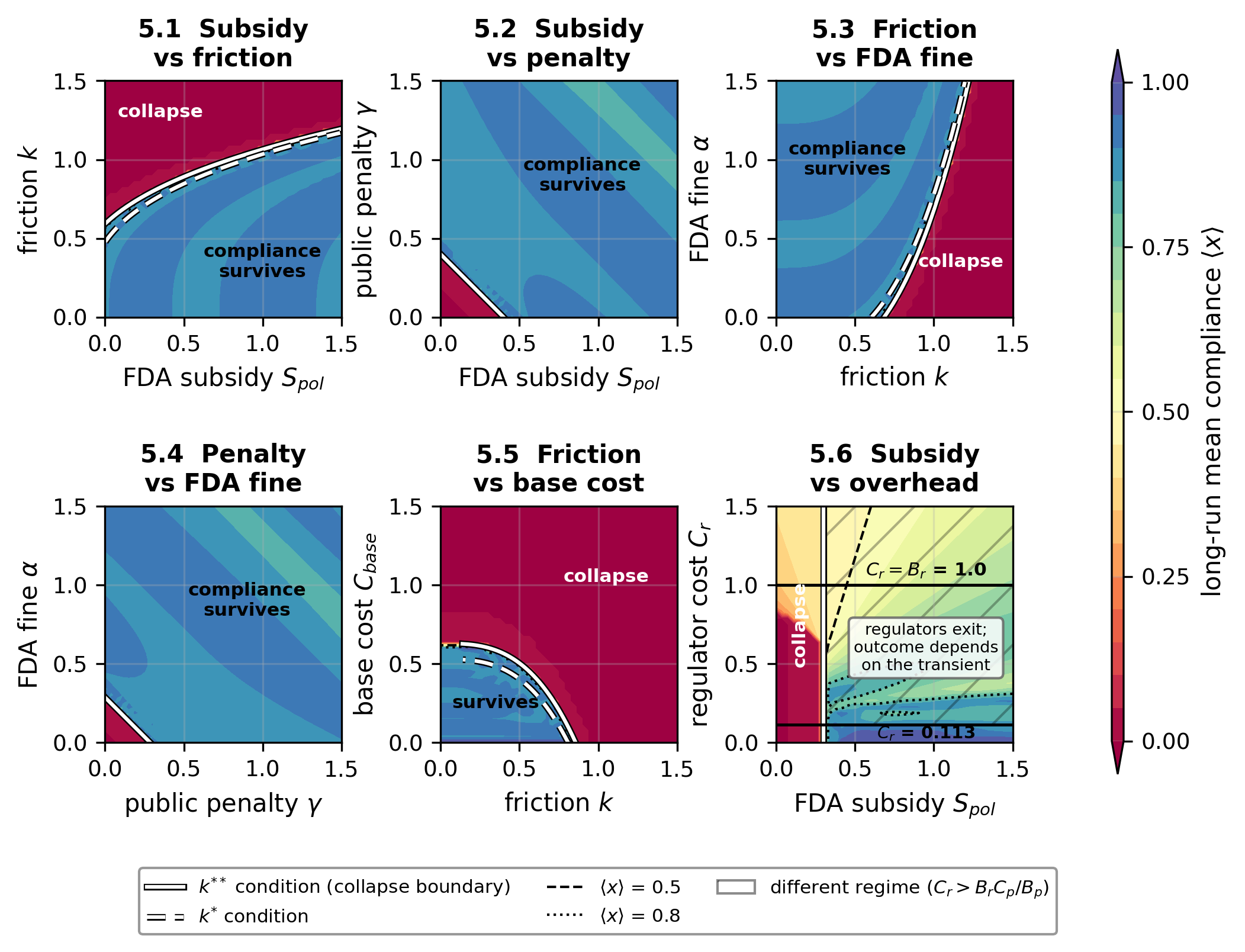}
    \caption{\textbf{Two-parameter maps of long-run compliance:} Every axis, including $C_r$ in 5.6, runs from $0$ to $1.5$, wide enough to contain the fitted values and both thresholds, and the panels are square so that all six maps share one scale. Each pixel of the $41\times41$ grid is one deterministic run (no noise) from $(0.5,0.5,0.5)$, colored by the long-run mean compliance $\langle x\rangle$ of \cref{eq:meanx} (red: collapse; blue: high compliance). Parameters not on an axis are held at their fitted values and $k=\kbar=0.7975$. One color bar (right) applies to all six panels, banded in steps of $0.05$; the black dashed and dotted lines are the $\langle x\rangle=0.5$ and $0.8$ contours. White solid (black-edged): the $k^{**}$ condition $S_{pol}+\alpha+\gamma=k^3+C_{base}$; white dashed (black-edged): the $k^*$ condition $S_{pol}+\alpha=k^3+C_{base}$. (5.1) subsidy vs friction; (5.2) subsidy vs public penalty; (5.3) friction vs fine; (5.4) public penalty vs fine; (5.5) friction vs base cost; (5.6) subsidy vs regulator cost, with the vertical line $S_{pol}=S^{**}$, the black lines $C_r=B_rC_p/B_p=0.113$ and $C_r=B_r=1.0$, and the regime above $0.113$ hatched (see text). Text labels name the regions.}\label{fig:heatmaps}
\end{figure}
\textbf{Analysis of \cref{fig:heatmaps} (parameter maps):} Each panel shows the long-run mean compliance over a pair of parameters, with the other constants at their fitted values. Since the mean averages over the cycle, the colors do not depend on where in the cycle a run happens to end. Two colors dominate each of the panels. Where compliance survives, its long-run mean is high (mostly between $0.8$ and $1$) and the panel is blue, because the time average of $x$ over a closed orbit is pinned near $x_z^*$ or $x_y^*$ (\cref{fig:verif_basin}). Where it does not, the panel is dark red. The boundary between the two is the information the maps carry, and in every panel where it applies that boundary is the analytic $k^{**}$ condition (white solid curve): compliance survives when the combined incentive $S_{pol}+\alpha+\gamma$ exceeds $k^3+C_{base}$.

\textbf{Panel 5.1} shows the subsidy needed to avoid collapse growing like $k^3$: compliance survives on $63\%$ of the map. The dashed $k^*$ curve lies inside the surviving region, because crossing $k^*$ changes which population enforces compliance, not whether compliance survives (\cref{fig:verif_bif}). \textbf{Panel 5.2} shows that at $\kbar$ compliance survives almost everywhere ($96\%$ of the map); collapse needs both a small subsidy and a small public penalty, $S_{pol}+\gamma<0.40$, so a weak public penalty is harmful only when the subsidy is also low. \textbf{Panel 5.3} shows that the fine needed to sustain compliance also grows like $k^3$: to the right of the white curve ($34\%$ of the map) compliance collapses at any fine in the plotted range. \textbf{Panel 5.4} shows public penalties and fines acting as substitutes: collapse occurs only when $\alpha+\gamma<0.29$ ($2\%$ of the map). \textbf{Panel 5.5} shows rising base costs compounding red tape. Compliance survives only in the lower-left $18\%$ of the map, bounded by $k=(S_{pol}+\alpha+\gamma-C_{base})^{1/3}$, so every increase in $C_{base}$ lowers the friction the market can tolerate. \textbf{Panel 5.6} maps the subsidy against regulator overhead $C_r$ on the same $0$--$1.5$ range as the other panels; the fitted value, $C_r=0.10$, lies at the bottom of the panel, just below the first of two lines that divide it. Below $C_r=B_rC_p/B_p=0.113$ (lower black line) the regulators' cost-to-benefit ratio is smaller than the public's and the market is in the fitted regime: collapse occurs for $S_{pol}<S^{**}=0.30$ (vertical line), because $C_r$ does not enter the $k^{**}$ condition, and above that subsidy the long-run mean is $0.85$--$1.00$. Above $0.113$ (hatched) the ratio is larger, so the sign of $\dot V$ in \cref{eq:V} reverses: regulators withdraw toward $y=0$ while public scrutiny rises toward $z=1$. Under the fitted equality $\gamma=C_{base}$ the payoff advantage $F$ then vanishes, so compliance freezes at whatever level it has reached when that happens. The long-run mean there ($0.47$--$0.92$) therefore depends on the transient rather than smoothly on the parameters, which produces the streaks in the hatched region. Above the upper black line, $C_r=B_r=1.0$, an inspection costs more than it can recover even in a market of defectors, so regulators leave whatever the state of the market. Compliance then stays near its starting value of $0.5$ ($0.46$--$0.70$), and for $C_r$ above about $0.9$ even subsidies below $S^{**}$ no longer produce a collapse. High regulator overhead thus does not by itself cause collapse, but it changes which population enforces compliance and makes the outcome depend on the market's history. The maps thus locate the parameter boundaries beyond which compliance collapses and show that they are given analytically by the friction threshold $k^{**}$.

\begin{figure}[H]
    \centering
    \includegraphics[width=\textwidth]{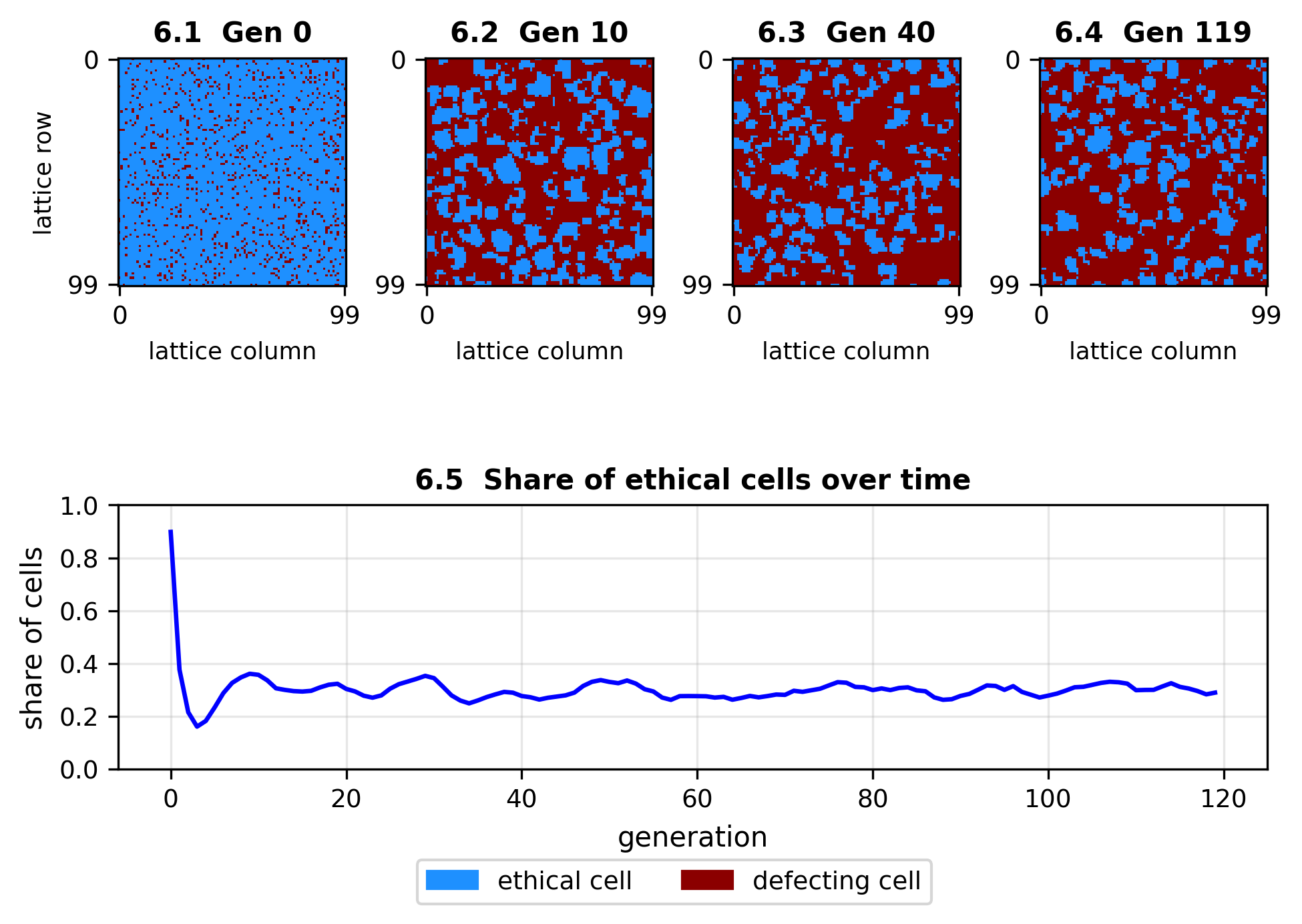}
    \caption{\textbf{Spatial cellular automaton:} A $100\times100$ lattice with periodic boundaries, Moore neighborhoods, and synchronous imitate-the-best updating; ethical cells earn $1$ and defecting cells $b=1.65$ per ethical neighbor; initially $10\%$ of cells, chosen at random, defect (seed 7). (6.1--6.4) the lattice at generations $0$, $10$, $40$, and $119$ (``Gen'' in the panel titles), with rows and columns indexed $0$--$99$ (blue: ethical cell; dark red: defecting cell); (6.5) share of ethical cells over $120$ generations, enough for the share to settle (it is flat after about generation $40$).}\label{fig:spatial}
\end{figure}
\textbf{Analysis of \cref{fig:spatial} (spatial dynamics):} Standard EGT assumes a perfectly mixed population, whereas spatial games capture local interactions. \textbf{Panels 6.1--6.4} show defectors invading their surroundings: isolated ethical cells are rapidly replaced, and ethical cells persist only in compact clusters, whose members mostly interact with one another. \textbf{Panel 6.5} shows the ethical share falling from $0.90$ to $0.36$ within ten generations and then fluctuating around $0.29$. In a well-mixed prisoner's dilemma with these payoffs defection would take over entirely; on the lattice, spatial reciprocity sustains a minority of cooperators \cite{nowak1992}. The analogy to geographic clustering of firms (for example the biotech hub around Kendall Square in Cambridge, Massachusetts) is suggestive rather than tested: this extension is illustrative and does not use the fitted constants.

\begin{figure}[H]
    \centering
    \includegraphics[width=0.8\textwidth]{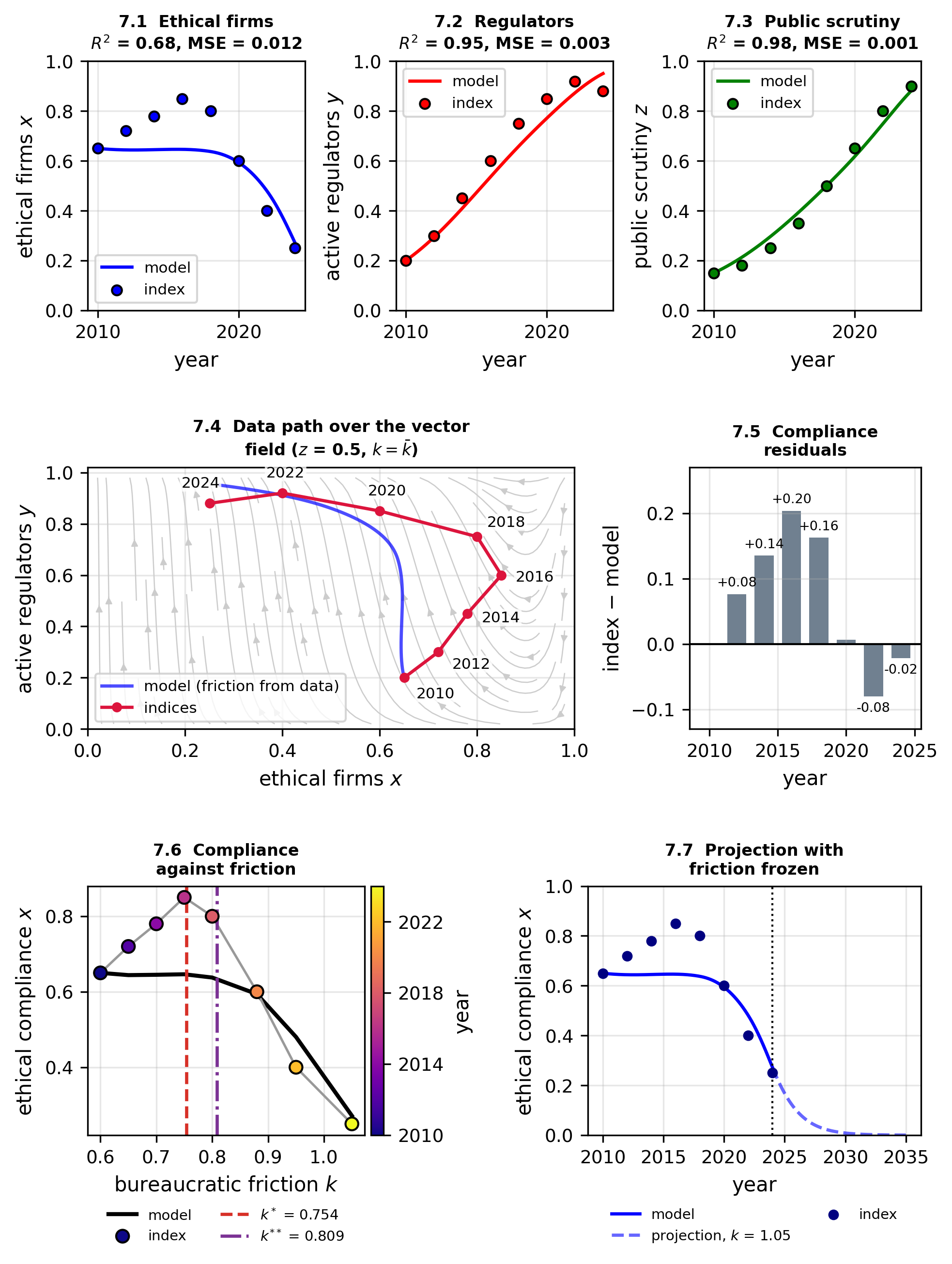}
    \caption{\textbf{Empirical calibration and projection:} (7.1--7.3) the calibrated model (lines; L-BFGS-B fit of \cref{sec:ml} with the empirical $k(t)$ as forcing) against the composite indices (points), with the $R^2$ and mean squared error (MSE) over the eight observation years given in each panel title and the same $0$--$1$ vertical scale in all three; (7.4) the indices and the calibrated path over the vector field $(\dot x,\dot y)$ of the constant-$k$ model, sliced at $z=0.5$, $k=\kbar$; (7.5) residuals $x_{\text{index}}-x_{\text{model}}$; (7.6) compliance against friction for the indices (colored by year; see the color bar) and the calibrated model at the observation years, with $k^*$ and $k^{**}$; (7.7) the calibrated path continued from its 2024 state with $k$ held at the 2024 value $1.05$, to 2035.}\label{fig:calib}
\end{figure}
\textbf{Analysis of \cref{fig:calib} (empirical calibration):} \textbf{Panels 7.1--7.3} compare the calibrated model with the composite indices; the fit is good for regulators ($R^2=0.95$) and the public ($R^2=0.98$) and moderate for compliance ($R^2=0.68$). The compliance fit misses the rise between 2012 and 2018 (residuals up to $+0.20$ in 2016, \textbf{Panel 7.5}). The reason is visible in Panel~8.2: along the empirical path the fitted payoffs give compliance essentially no fitness advantage between 2010 and 2016 ($f_1^E-f_1^D$ between $-0.022$ and $+0.009$), so the model cannot generate a sustained rise there, and the objective's heavier weighting of 2020--2024 favors fitting the decline instead. The early rise must therefore come from factors the model holds constant, such as changing fines, subsidies, or enforcement priorities, or from exogenous shocks. \textbf{Panel 7.4} overlays the indices and the calibrated path on the model's vector field; the indices depart from the calibrated path in the middle of the period (2016--2018) and rejoin it from 2020 on. Since the fit is deliberately anchored to the most recent, most volatile years, we present the calibrated model as a starting point for empirical EGT modeling of the industry rather than as a definitive fit. \textbf{Panel 7.6} plots compliance against friction. The compliance index peaks at $k=0.75$ (2016), within rounding of the computed $k^*=0.754$, falls slightly by $k=0.80$ (2018), and then falls steeply after friction passes $k^{**}=0.809$ ($0.60$ at $k=0.88$ in 2020, $0.25$ at $k=1.05$ in 2024); the calibrated path is nearly flat until $k\approx0.80$ and declines thereafter. The computed thresholds and the empirical friction values are reported separately throughout: $0.88$ is the index value in 2020, not a computed threshold. Finally, \textbf{Panel 7.7} continues the calibrated path from 2024 with friction held at its 2024 value, which lies beyond $k^{**}$: compliance continues to decline, to about $0.01$ by 2030. This projection assumes no change in any parameter and should be read as the consequence of current settings, not as a forecast.

\begin{figure}[H]
    \centering
    \includegraphics[width=\textwidth]{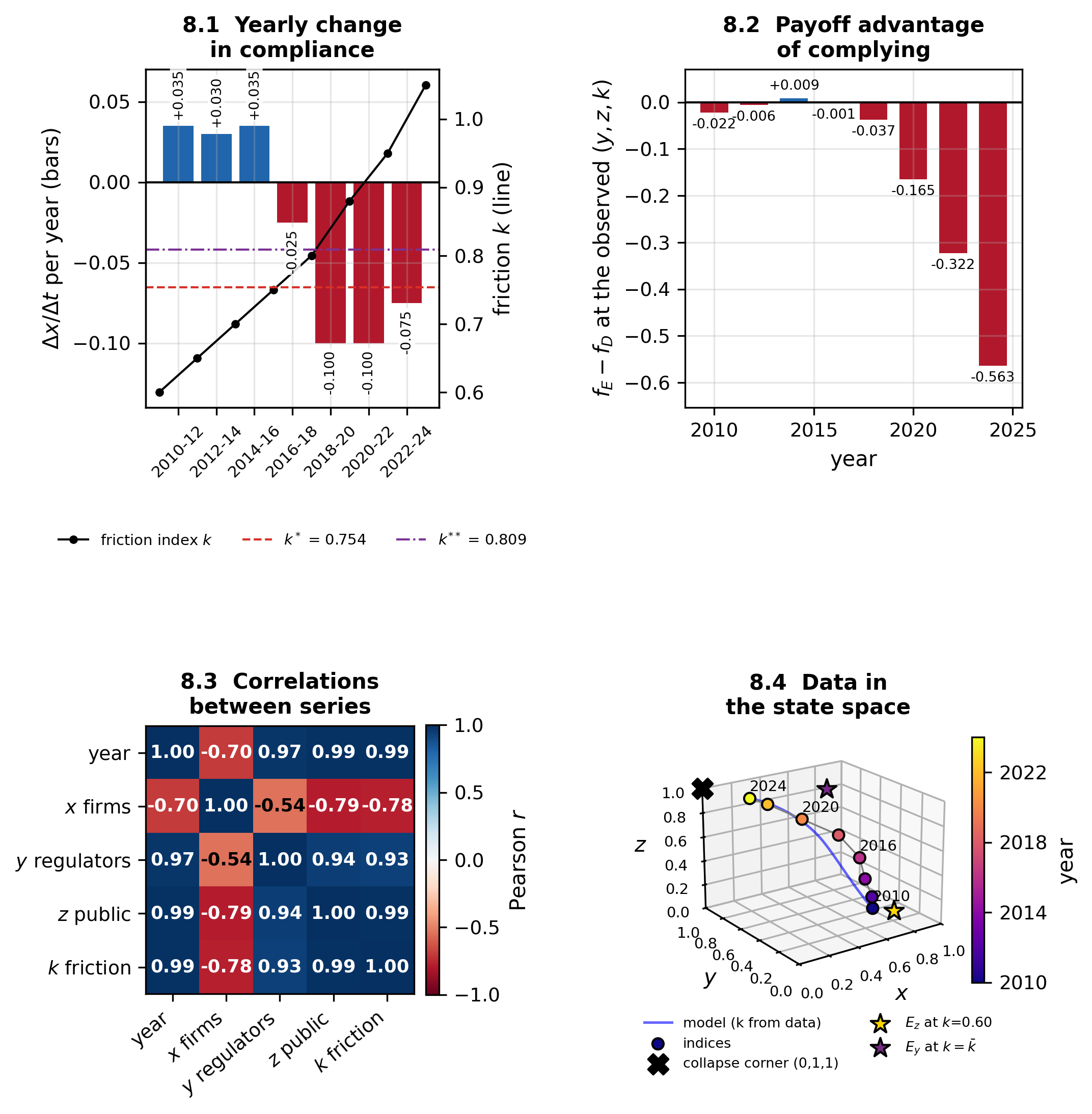}
    \caption{\textbf{Synthesis of the empirical record:} (8.1) biennial rate of change $\Delta x/\Delta t$ of the compliance index (bars, values printed) with the friction index (line, right axis) and the thresholds $k^*$ (red dashed) and $k^{**}$ (purple dash-dotted), keyed in the legend below the panel; (8.2) the fitness advantage of compliance, \cref{eq:fE}, evaluated with the fitted constants at the observed $(y_t,z_t,k_t)$; (8.3) Pearson correlations among the year and the four indices (\texttt{Evolutionary\_t}, which is the year minus 2010, is omitted; blue for positive and red for negative correlation, with white labels on the strongly correlated cells); (8.4) the indices in the state space $(x,y,z)$ (colored by year; see the color bar), the calibrated path, the collapse corner $(0,1,1)$, $E_z$ at $k=0.60$, and $E_y$ at $k=\kbar$.}\label{fig:topo}
\end{figure}
\textbf{Analysis of \cref{fig:topo} (synthesis):} \textbf{Panel 8.1} shows how fast compliance changed. The index rose by $0.030$--$0.035$ per year from 2010 to 2016, fell by $0.025$ per year between 2016 and 2018 (the interval in which friction crossed $k^*$), and fell four times as fast, $0.10$ per year, from 2018 to 2022, after friction exceeded $k^{**}$; the decline slowed slightly to $0.075$ per year in 2022--2024. \textbf{Panel 8.2} evaluates the fitted payoffs along the observed path. The fitness advantage of compliance is close to zero from 2010 to 2016, turns negative in 2018 ($-0.037$), and then falls to $-0.165$ (2020), $-0.32$ (2022), and $-0.56$ (2024): as friction rises, the fitted payoffs penalize compliance increasingly strongly. \textbf{Panel 8.3} shows the empirical correlations. Compliance is negatively correlated with friction ($r=-0.78$) and with public scrutiny ($r=-0.79$), and less strongly with regulator activity ($r=-0.54$), while regulators, the public, and friction rise together ($r\ge0.93$). With eight observations sharing common trends these correlations describe co-movement, not causation. \textbf{Panel 8.4} places the empirical record in the model's state space. The indices move from near the firm--regulator cycle around $E_z$ in 2010 toward the corner $(0,1,1)$, the model's collapse state for $k>k^{**}$ (no ethical firms, regulators and the public fully mobilized); the distance from that corner falls from $1.34$ in 2010 to $0.30$ in 2024. The industry did not decline gradually in this picture: its path turned toward the collapse corner once friction passed the computed thresholds.

\subsection{Model Validation: Fixed Points, Conserved Quantity, Long-Run Behavior}\label{sec:validation}
To confirm that the qualitative claims above are produced by the calibrated system rather than by a fixed integration horizon, we integrated the constant-$k$ model to long times ($t$ up to $1700$) and checked each analytic result of \cref{sec:jac,sec:longrun} numerically.

\begin{figure}[H]
    \centering
    \includegraphics[width=0.9\textwidth]{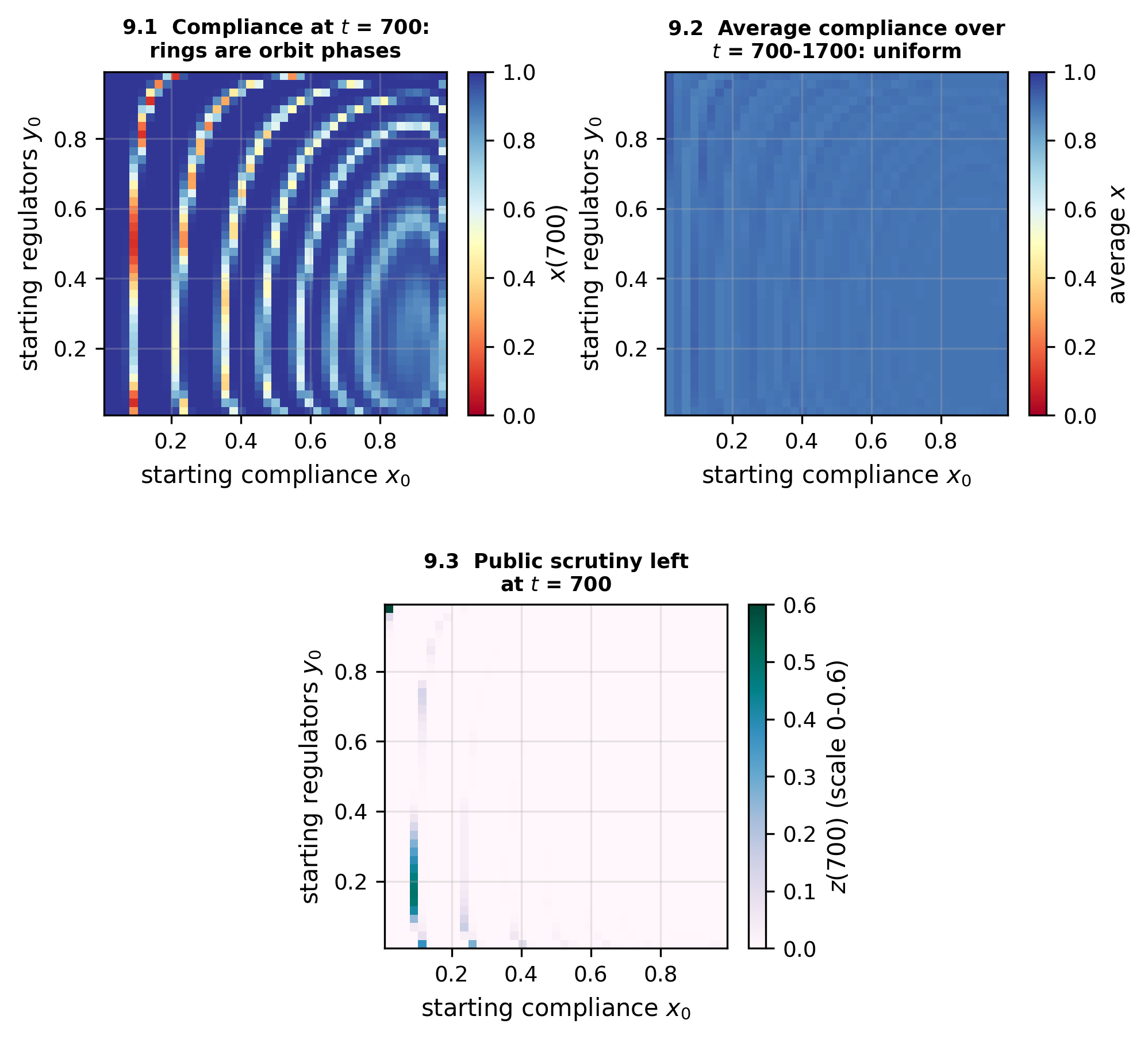}
    \caption{\textbf{Long-run behavior from a grid of initial conditions} ($41\times41$ values of $(x_0,y_0)\in[0.02,0.98]^2$, $z_0=0.4$, $k=0.60$). (9.1) compliance at $t=700$; (9.2) mean compliance over $700\le t\le1700$; (9.3) public scrutiny at $t=700$. Panels 9.1 and 9.2 share a $0$--$1$ color scale; Panel 9.3 uses $0$--$0.6$, the largest value of $z$ that remains.}\label{fig:verif_basin}
\end{figure}
\paragraph{The rings in Panel 9.1 are orbits, not basins:} \Cref{fig:verif_basin} starts the model from $41\times41$ different combinations of initial compliance $x_0$ (horizontal axis) and initial regulator activity $y_0$ (vertical axis), all with $z_0=0.4$ and $k=0.60$, and colors each starting point by what the run looks like much later. \textbf{Panel 9.1} colors it by compliance at the single time $t=700$ and shows concentric rings of high and low compliance. The rings are not basins of attraction. Each starting point ends up on its own closed orbit (\cref{fig:basins}), and compliance at a fixed time only records where on that orbit the run happens to be at that moment; neighboring starting points sit at slightly different phases, which produces the rings. \textbf{Panel 9.2} averages compliance over $700\le t\le1700$ instead, which covers more than six cycles and so removes the phase. The rings disappear: the average lies between $0.87$ and $0.94$ for every starting point, close to $x_z^*=0.90$. This is the Lotka--Volterra time-average identity: over a closed orbit on $z=0$, $\ln[y/(1-y)]$ returns to its starting value, so the time average of $G=b_0+b_x x$ vanishes and $\langle x\rangle=x_z^*$ \cite{hofbauer1998}. \textbf{Panel 9.3} shows how much public scrutiny is left at $t=700$. It has fallen below $0.05$ for $98.2\%$ of starting points. The exceptions start with few ethical firms ($x_0\approx0.1$), and their orbits pass close to the corners of the cube, where the dynamics are slow.

\begin{figure}[H]
    \centering
    \includegraphics[width=0.9\textwidth]{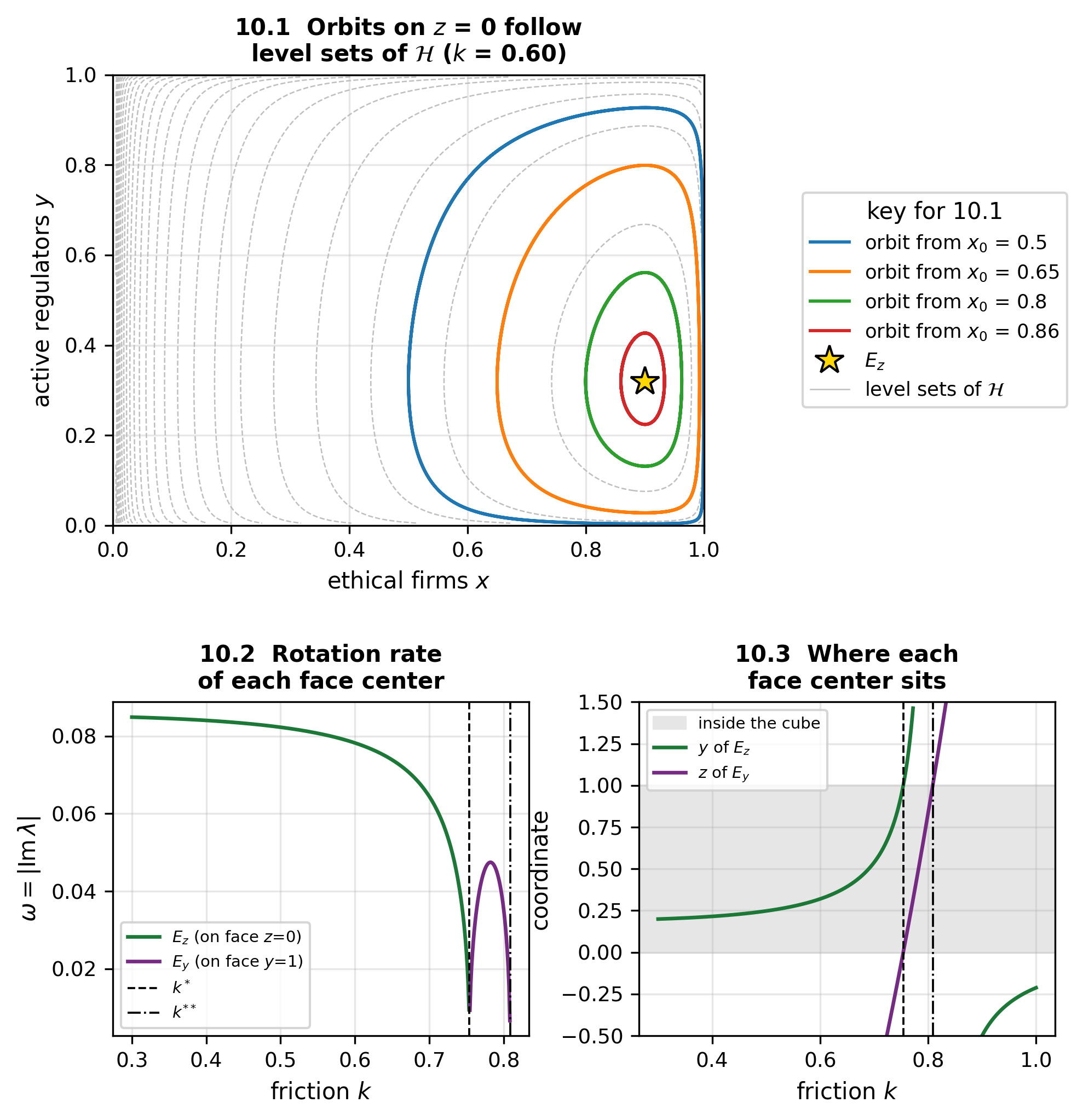}
    \caption{\textbf{Conserved quantity and stability of the face centers:} (10.1) on the face $z=0$ at $k=0.60$: level sets of $\mathcal{H}$ (\cref{eq:H}; gray) and four trajectories from $(x_0,0.3,0)$, colored by $x_0$, integrated to $t=300$; gold star: $E_z$. (10.2) in-face frequency $\omega=|\mathrm{Im}\,\lambda_\parallel|$ of $E_z$ (green) and $E_y$ (purple) against $k$, shown where each lies in the cube. (10.3) the $y$-coordinate of $E_z$ and the $z$-coordinate of $E_y$ against $k$; the gray band is $[0,1]$. Dashed: $k^*=0.754$; dash-dotted: $k^{**}=0.809$.}\label{fig:verif_stab}
\end{figure}
\paragraph{The face fixed points are true centers:} \textbf{Panel 10.1} of \cref{fig:verif_stab} is the face $z=0$ at $k=0.60$. The gray curves are level sets of the conserved quantity $\mathcal{H}$ of \cref{eq:H}, computed from the formula alone; the colored curves are four simulated orbits started at different compliance levels. The simulated orbits lie exactly on the level sets: along a $z=0$ orbit integrated to $t=1000$, $\ln\mathcal{H}$ varies by a relative amount of $6\times10^{-9}$, which is the integration tolerance. Since the orbits are level sets of a conserved quantity, they are closed and neither spiral in nor out, so $E_z$ is a true center rather than a weak spiral. Panels 10.2 and 10.3 follow the two face centers as friction changes. \textbf{Panel 10.2} plots how fast the market rotates around each center (the imaginary part of its eigenvalues). The rotation rate of $E_z$ falls to zero as $k$ approaches $k^*$, where $E_z$ becomes a saddle, and $E_y$ rotates only for $k^*<k<k^{**}$. \textbf{Panel 10.3} shows why, by plotting where each center sits. The $y$-coordinate of $E_z$ (green) leaves the cube through $y=1$ at $k^*$. At that point the $z$-coordinate of $E_y$ (purple) enters the cube through $z=0$, and it leaves through $z=1$ at $k^{**}$. For every $k<k^{**}$ exactly one face center lies inside the cube, and no eigenvalue crosses the imaginary axis away from zero, so no Hopf bifurcation occurs.

\begin{figure}[H]
    \centering
    \includegraphics[width=\textwidth]{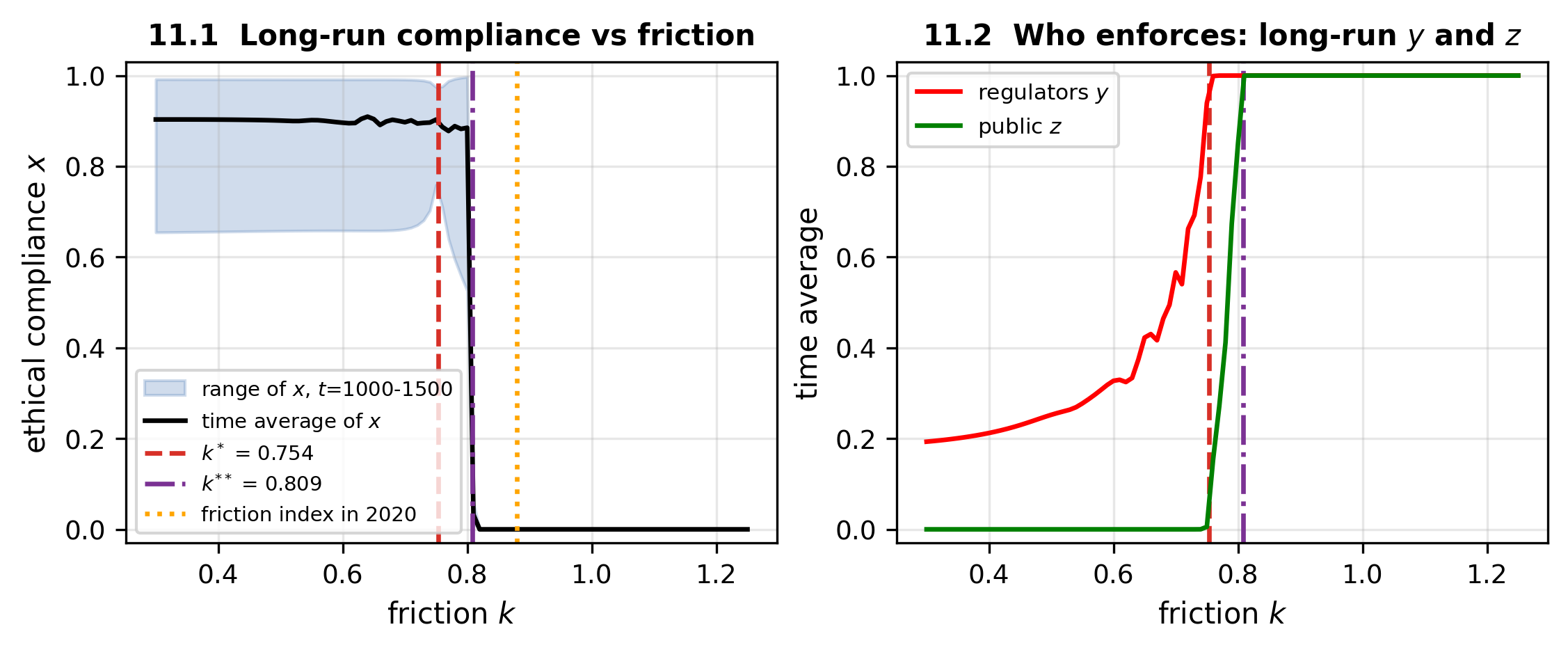}
    \caption{\textbf{Bifurcation diagram in the friction $k$.} $96$ values of $k\in[0.3,1.25]$, each integrated from $(0.65,0.20,0.15)$ with statistics over $1000\le t\le1500$. (11.1) min--max range (band) and time mean (line) of compliance $x$; (11.2) time means of regulators $y$ and public scrutiny $z$. Dashed: $k^*$; dash-dotted: $k^{**}$; dotted: the friction index in 2020 ($k=0.88$).}\label{fig:verif_bif}
\end{figure}
\paragraph{Two thresholds, and a hand-over of enforcement:} \Cref{fig:verif_bif} sweeps friction from $0.3$ to $1.25$ in steps of $0.01$, runs the model from the 2010 state for each value, and summarizes the run over $1000\le t\le1500$, long after the start-up transient. In \textbf{Panel 11.1} the shaded band is the range compliance covers during the cycle and the black line its time average. The time average stays near $0.9$ for every $k<k^{**}$ and drops to zero beyond it, while the band narrows as $k$ approaches $k^{**}$. \textbf{Panel 11.2} shows what happens between the thresholds. Mean regulator activity rises with friction and saturates at $1$ at $k^*$; mean public scrutiny is zero below $k^*$ and rises to $1$ at $k^{**}$. Below $k^*$ compliance is enforced by regulators alone; between $k^*$ and $k^{**}$ regulators are fully mobilized and public scrutiny supplies the remaining incentive; beyond $k^{**}$ even full mobilization of both cannot offset friction, and the system converges to $(0,1,1)$. The friction index passed $k^*$ shortly after 2016, passed $k^{**}$ between 2018 and 2020, and stood at $0.88$ in 2020. This ordering, together with the rise of the regulator and public indices over the same period, is consistent with the hand-over shown in Panel 11.2.

\begin{figure}[H]
    \centering
    \includegraphics[width=0.95\textwidth]{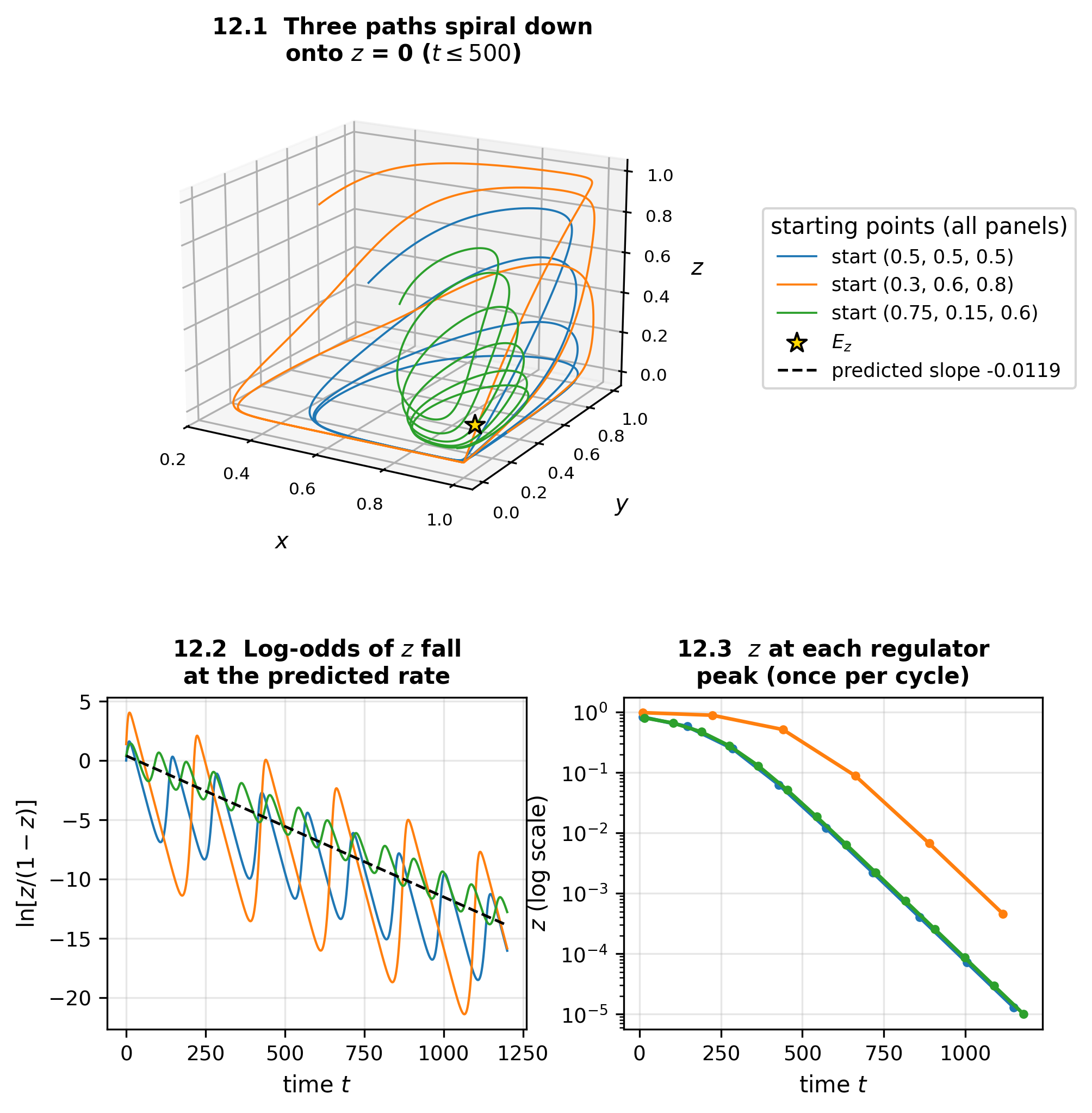}
    \caption{\textbf{Approach to the face $z=0$} at $k=0.60$, from $(0.5,0.5,0.5)$, $(0.3,0.6,0.8)$, and $(0.75,0.15,0.6)$. (12.1) the trajectories for $t\le500$; gold star: $E_z$. (12.2) log-odds of public scrutiny, $\ln[z/(1-z)]$, against $t$, with the slope $B_p(x_y^*-x_z^*)=-0.0119$ predicted by \cref{eq:V} (dashed). (12.3) $z$ at successive maxima of $y$ (one point per cycle), on a logarithmic scale; colors in all panels match the starts listed in (12.1). The key at the top right applies to all three panels.}\label{fig:zdecay}
\end{figure}
\paragraph{Public scrutiny decreases every cycle:} \Cref{fig:zdecay} follows three trajectories at $k=0.60$, started from the points listed in the key panel. \textbf{Panel 12.1} shows them in the cube for $t\le500$: each spirals around $E_z$ while sinking toward the floor $z=0$. \textbf{Panel 12.2} plots the log-odds of public scrutiny, $\ln[z/(1-z)]$, which \cref{eq:V} predicts should fall on average along a straight line of slope $B_p(x_y^*-x_z^*)=-0.0119$ (dashed). The curves oscillate with the cycle but follow that line. \textbf{Panel 12.3} keeps one value of $z$ per cycle (at each peak of regulator activity) on a logarithmic axis, so a constant rate of decay appears as a straight line. After a short transient each sequence falls geometrically. For the trajectory of \cref{fig:core}, the ratio between successive peaks settles at $0.288$, matching the prediction $e^{-0.0119\times104.6}=0.288$ for its period of $104.6$ time units.

\paragraph{No evidence of chaos:} \textbf{Panel~13.1} shows the estimates converging slowly: two exponents approach zero and the third approaches the transverse rate $-0.0119$. Short averaging times give misleading values (at $T=10$ the estimates are $0.0044$, $-0.1016$, $-0.1427$). In \textbf{Panel 13.2}, $\lambda_1$ lies between $-0.0014$ and $0.0021$ for every run and $\lambda_3$ between $-0.019$ and $-0.007$. A largest exponent of this size, together with a second exponent of the same order and the vanishing exponent required along the flow \cite{haken1983}, is the signature of neutral, non-chaotic motion.

\paragraph{Summary:} All checks agree with the analytic treatment of \cref{sec:jac,sec:longrun}: the dynamics approach the boundary at the rate set by \cref{eq:V}; the face fixed points are true centers with a conserved quantity; compliance survives in the long run exactly while $k<k^{**}$; enforcement passes from regulators to the public at $k^*$; and there is no sign of chaos. An alternative modeling route, re-parameterizing so that public scrutiny survives in the long run or adding a weak stabilizing term that converts the centers into genuine attractors, is left as future work.

\subsection{External Empirical Context}\label{sec:empirical}
The following figures are compiled from public sources and do not depend on the ODEs; they place the model's mechanisms in their real-world setting.

\begin{figure}[H]
    \centering
    \begin{minipage}[t]{0.49\textwidth}
        \centering
        \includegraphics[width=\textwidth]{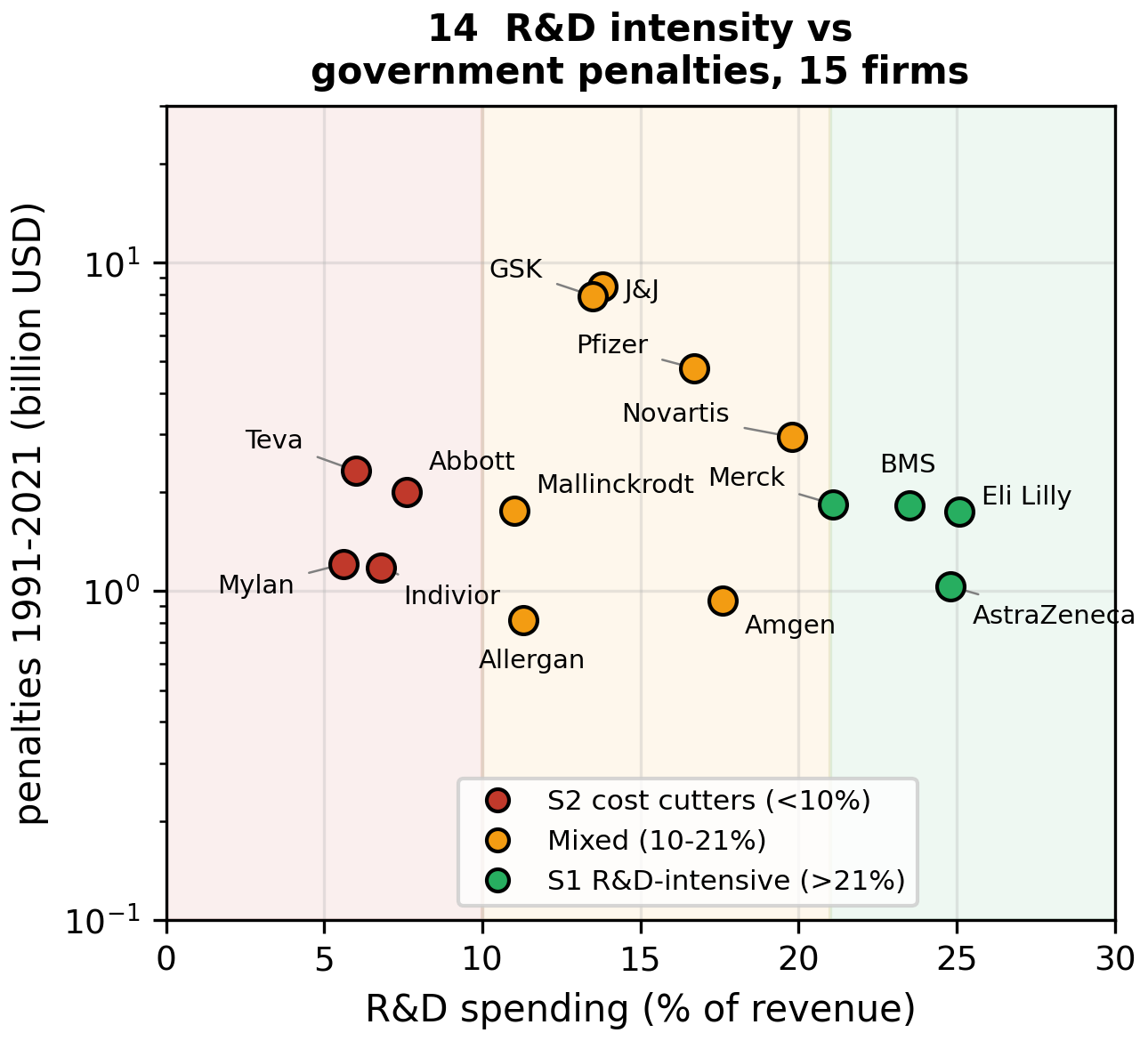}
        \captionsetup{justification=raggedright,singlelinecheck=false}
        \caption{R\&D investment (\% of revenue) against government penalties over 1991--2021 (logarithmic scale) for the 15 firms of \cref{tab:full_firms}, colored by strategy zone (shaded bands: zone thresholds). Penalty bases and sources are listed in Appendix~\ref{app:firms}.}\label{fig:rnd_penalty}
    \end{minipage}\hfill
    \begin{minipage}[t]{0.49\textwidth}
        \centering
        \includegraphics[width=\textwidth]{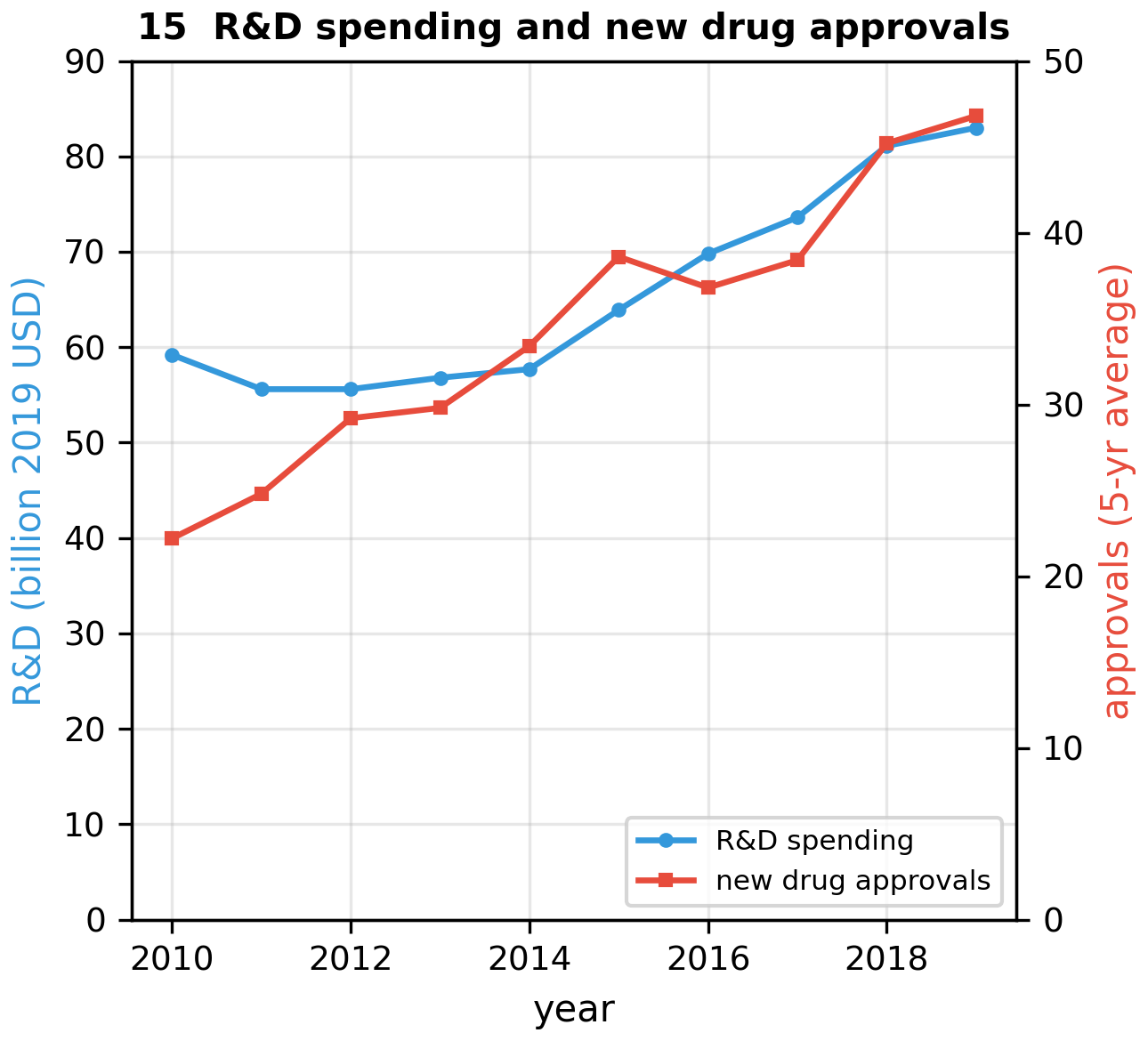}
        \captionsetup{justification=raggedright,singlelinecheck=false}
        \caption{Pharmaceutical R\&D spending (billions of 2019 dollars, left) and new drug approvals (five-year moving average, right), 2010--2019; data from the Congressional Budget Office \cite{cbo2021}. Both axes start at zero, but are not scaled in the same manner.}\label{fig:rnd_approvals}
    \end{minipage}
\end{figure}
\begin{figure}[H]
    \centering
    \includegraphics[width=0.9\textwidth]{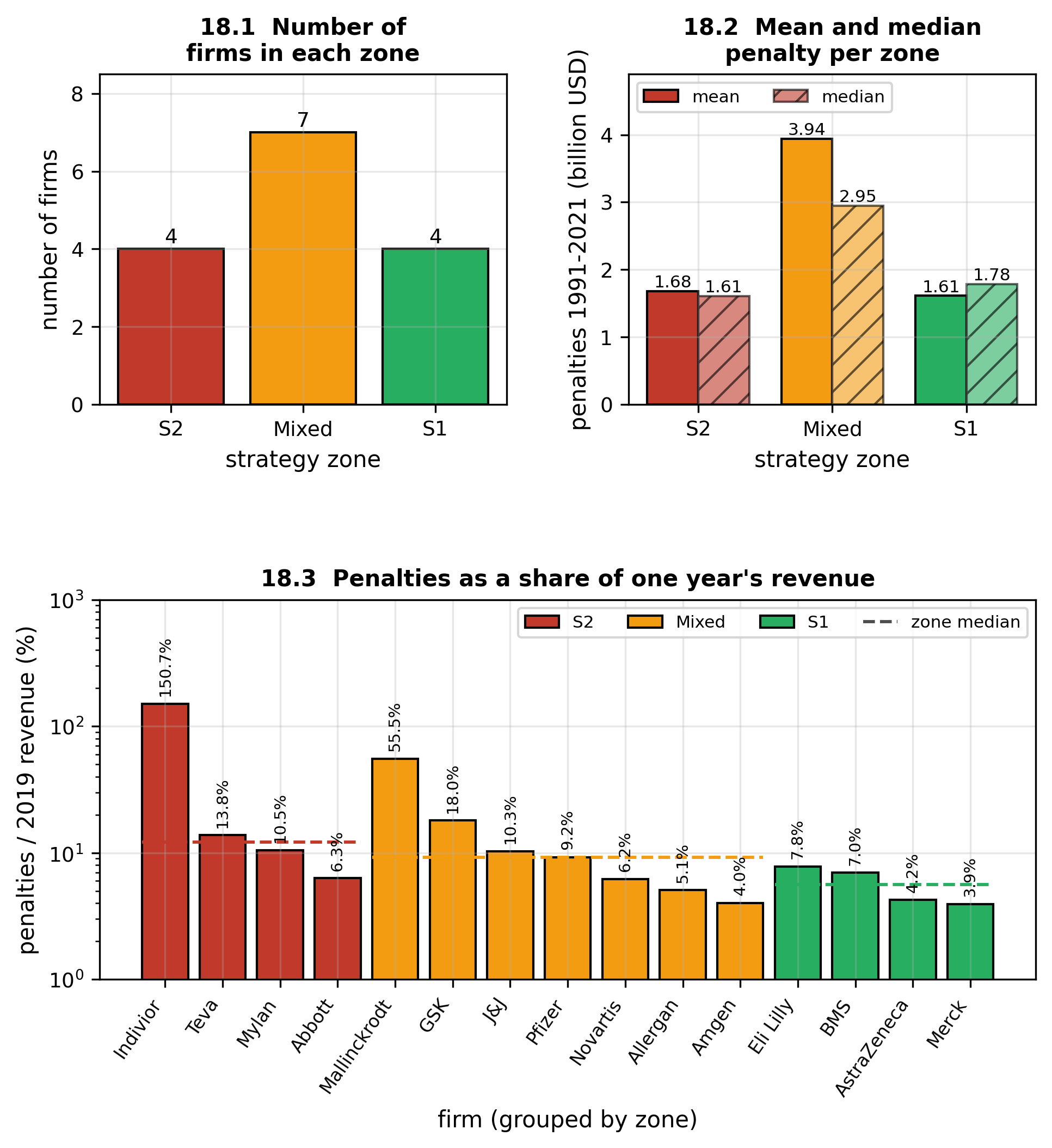}
    \caption{\textbf{Summary of the firms in \cref{fig:rnd_penalty} by strategy zone.} (18.1) number of firms per zone; (18.2) mean (solid) and median (hatched) government penalties over 1991--2021 per zone; (18.3) each firm's penalties divided by its fiscal-2019 revenue (logarithmic scale; dashed lines: zone medians). Revenue sources are listed in Appendix~\ref{app:firms}.}\label{fig:zonehist}
\end{figure}

\textbf{Analysis of \cref{fig:rnd_penalty,fig:rnd_approvals} (strategy and innovation):} To compare the model's strategy types with real firms, we classified the 15 firms selected by the inclusion rule of \cref{sec:data} by R\&D intensity. \textbf{\Cref{fig:rnd_penalty}} shows that, in absolute dollars, penalties are not ordered by strategy zone. The two largest totals in the sample belong to Mixed firms (Johnson \& Johnson, \$8.4 billion, and GlaxoSmithKline, \$7.9 billion), while the S2 totals range from \$1.2 to \$2.3 billion and the S1 totals from \$1.0 to \$1.8 billion. Absolute penalties largely track firm size: the large diversified firms accumulate the biggest totals while remaining solvent. Penalties relative to revenue, the more informative comparison, are shown in \cref{fig:zonehist}. We stress that these are associations across fifteen firms, not identified causal effects. \textbf{\Cref{fig:rnd_approvals}} illustrates the prisoner's-dilemma race: from 2010 to 2019 R\&D spending rose from \$59 billion to \$83 billion per year (in 2019 dollars) while the moving average of new drug approvals roughly doubled \cite{cbo2021}. The two series move together, which is consistent with the first-mover pressure described above, but the figure documents co-movement rather than establishing that either caused the other.

\begin{figure}[H]
    \centering
    \begin{minipage}[t]{0.49\textwidth}
        \centering
        \includegraphics[width=\textwidth]{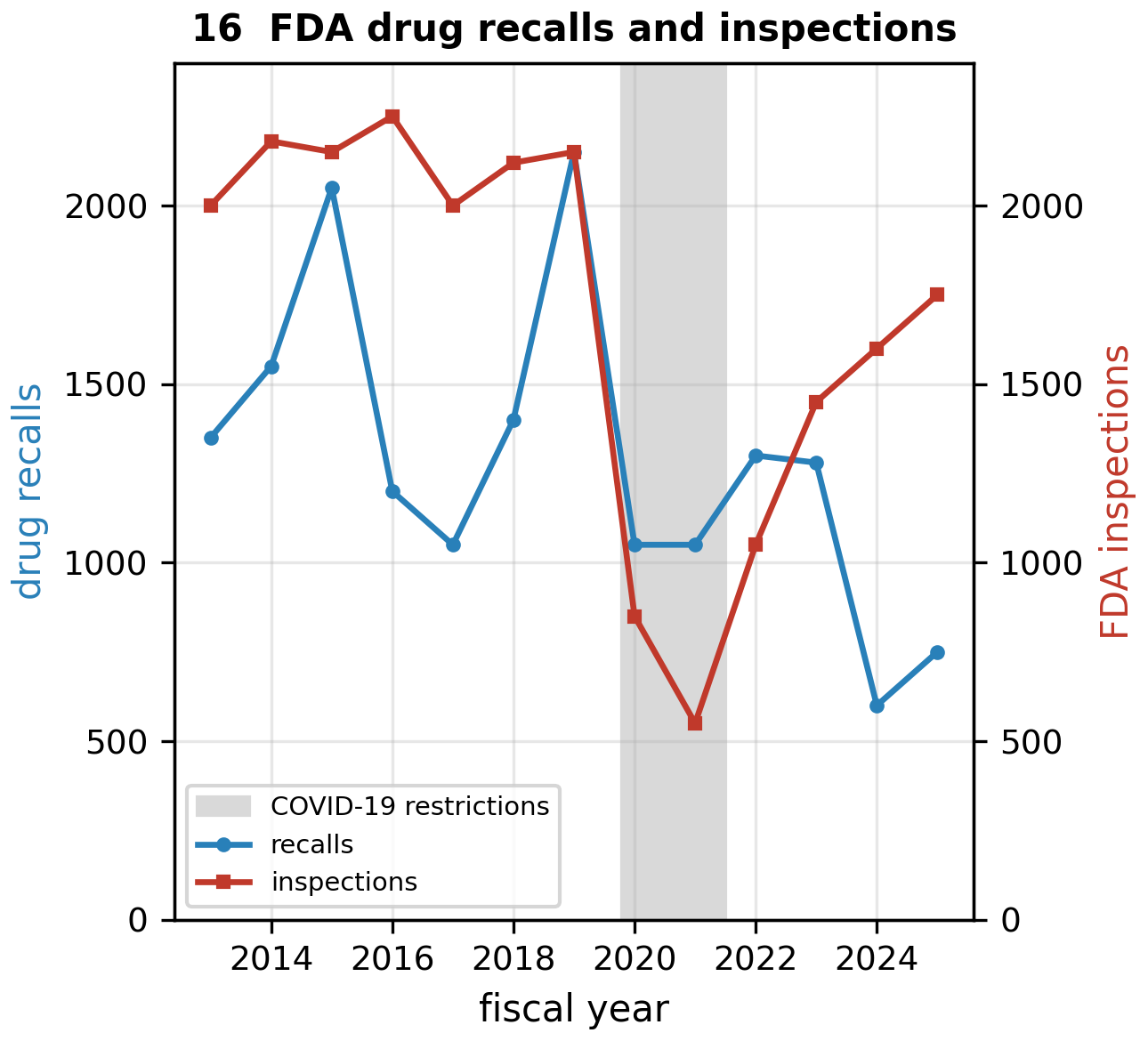}
        \captionsetup{justification=raggedright,singlelinecheck=false}
        \caption{FDA drug recalls (left axis) and FDA inspections (right axis), 2013--2025; the shaded band marks the period of COVID-19 travel and facility-access restrictions. Both axes start at zero, but are not scaled in the same manner. Data: FDA Data Dashboard \cite{fdadashboard}.
        }\label{fig:recalls}
    \end{minipage}\hfill
    \begin{minipage}[t]{0.49\textwidth}
        \centering
        \includegraphics[width=\textwidth]{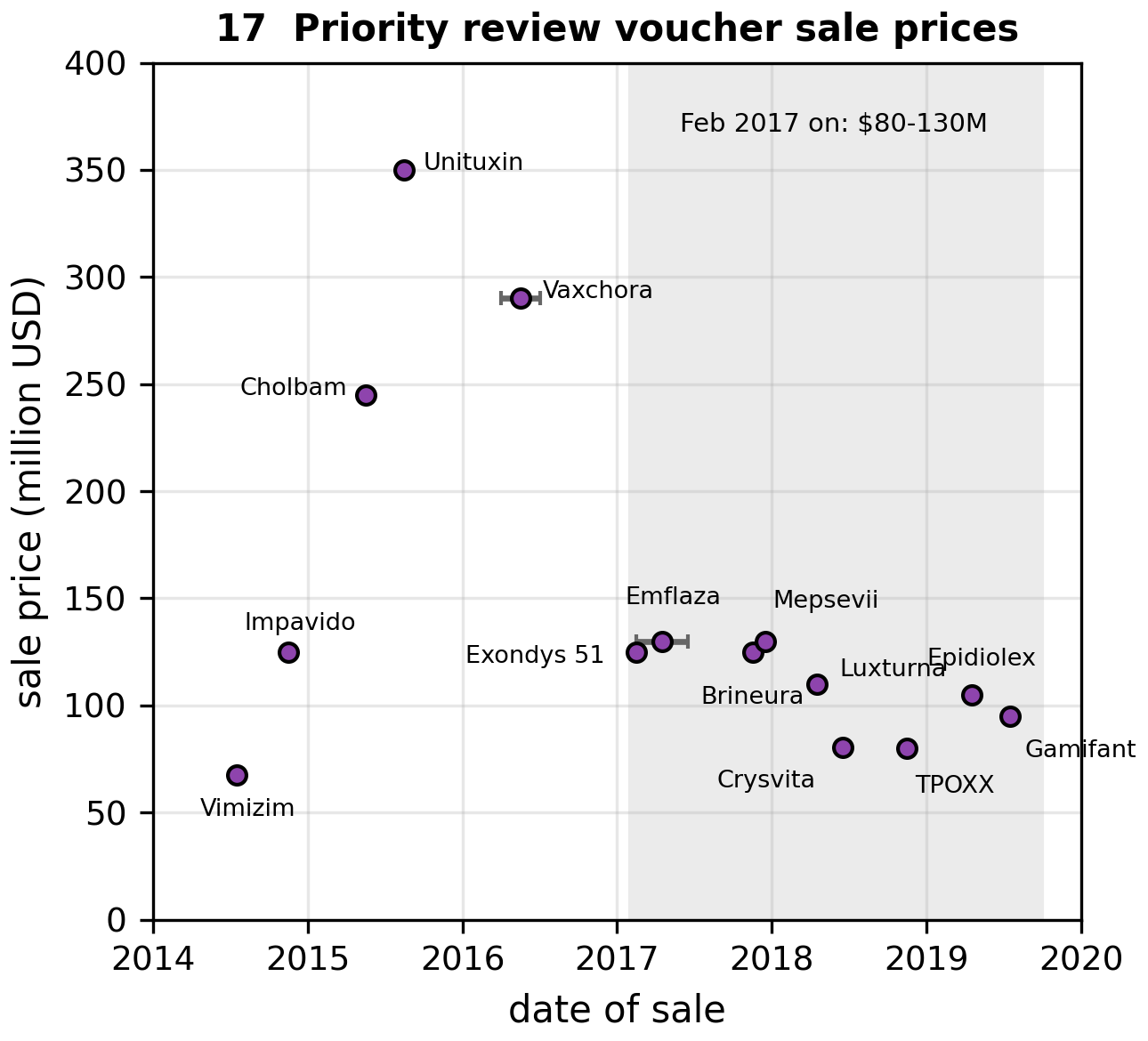}
        \captionsetup{justification=raggedright,singlelinecheck=false}
        \caption{Sale prices of the 14 priority review vouchers (PRVs) sold with a disclosed price through fiscal year 2019, plotted at their dates of sale \cite{gao2020}; horizontal bars mark the two sales dated only to a quarter or a range, and the shaded region marks sales from February 2017 on.
        }\label{fig:prv}
    \end{minipage}
\end{figure}
\textbf{Analysis of \cref{fig:recalls,fig:prv} (inspection games and the valley of death):} \textbf{\Cref{fig:recalls}} shows the inspection game in the regulatory record. FDA inspections dropped sharply between 2020 and 2021, a decline coinciding with COVID-19 travel and facility-access restrictions, which in the model raises the expected payoff of defection; recall activity stayed between about 1,050 and 1,300 per year over 2020--2023. The largest recall count in the series (2,150) occurs in 2019, before the inspection decline, so the 2020--2021 drop is better read as coinciding with a period of already-high recall activity than as its cause. \textbf{\Cref{fig:prv}} addresses the volunteer's dilemma through the secondary market for priority review vouchers \cite{ridley2006}. A PRV is a transferable entitlement, awarded by the FDA to the sponsor of a qualifying neglected-disease or rare-pediatric-disease drug, that grants its holder a shortened (priority) review for a future application and may be sold to another firm. In the language of our model a PRV is a non-punitive subsidy: it raises $S_{pol}$ for firms that undertake foundational research without requiring direct appropriation, and so acts on the volunteer's-dilemma payoff $U_v$ of \cref{sec:games}. According to the Government Accountability Office, 17 of the 31 vouchers awarded through fiscal year 2019 were sold, and prices are disclosed for 14 \cite{gao2020}. The first sales set the highest prices: \$67.5 million (Vimizim, July 2014), \$125 million (Impavido, November 2014), \$245 million (Cholbam, May 2015), and a peak of \$350 million (Unituxin, August 2015), followed by \$290 million (Vaxchora, second quarter of 2016). From February 2017 on, as more vouchers were awarded and offered for sale, every disclosed price fell between \$80 million and \$130 million. Over the same 2015--2017 window, recalls fell from their 2015 high of 2,050 to about 1,050 in 2017 (\cref{fig:recalls}); we present this as co-timing rather than as evidence that vouchers reduced recalls. What the two figures do show is that a subsidy instrument worth \$80--350 million per award exists and is actively traded, which is the order of magnitude that, in the model, moves $S_{pol}$ enough to matter near the thresholds (Panel~5.1). Such subsidies offer a way across the ``valley of death'' that does not rely on penalties.

To summarize the fifteen firms behind \cref{fig:rnd_penalty}, \cref{fig:zonehist} groups them by strategy zone. \textbf{Panel 18.1} shows the counts that the inclusion rule and the two cut-offs produce: $4$ S2, $7$ Mixed, and $4$ S1 firms. \textbf{Panel 18.2} shows that in absolute dollars the Mixed zone carries the most: its mean penalty is \$3.94 billion and its median \$2.95 billion, against \$1.68 and \$1.61 billion for S2 firms and \$1.61 and \$1.78 billion for S1 firms. Absolute totals mostly reflect firm size, because a larger firm sells more of the products a settlement is about. \textbf{Panel 18.3} removes that effect by dividing each firm's penalties by its revenue in a single year (2019). On that scale the ordering is the one the model suggests. The median ratio is $12.2\%$ of a year's revenue for S2 firms, $9.2\%$ for Mixed firms, and $5.6\%$ for S1 firms, and as shown in \cref{sec:data} this ordering survives the removal of any single firm. The two largest ratios belong to firms whose revenue is small relative to their settlements: Indivior, penalized more than one and a half years of its 2019 revenue, and Mallinckrodt, more than half a year's. No S1 firm's penalties reach $8\%$ of a year's revenue. Cost cutters are therefore penalized most heavily relative to their size, while large diversified firms carry the largest absolute penalties at a cost of a few percent of one year's revenue. This ordering is consistent with the model's association between low R\&D investment and defection, with the caveat that it rests on fifteen firms and on revenue in a single year.

\section{Discussion and Future Directions}\label{sec:discussion}
The identification of a ``regulatory paradox'' within our EGT framework provides a critique of a common assumption in public health policy. Traditional regulatory reasoning assumes a monotonic relationship: more regulatory oversight yields proportionally more ethical compliance. In our model this holds only up to a point. Oversight that raises the cost of complying faster than it raises the incentive to comply eventually collapses compliance (Panels~5.1 and 5.3 and \cref{fig:verif_bif}).

As bureaucratic friction $k$ increases, the cost of compliance $C_{base}+k^3 y$ outpaces the combined incentive of subsidies, fines, and public penalties. Beyond the threshold $k^{**}$, corporate actors in the model treat state fines as a cost of doing business rather than as a deterrent. Our collected dataset is consistent with this reading: Pfizer's \$2.3 billion settlement in 2009 was the largest health-care fraud settlement in U.S.\ history at the time \cite{doj2009}; yet even Pfizer's total government penalties over 1991--2021, \$4.8 billion \cite{publiccitizen2024}, amount to about $9\%$ of a single year's revenue (\cref{fig:zonehist}), too little to alter its payoff matrix sufficiently to prevent future defection.

\subsection{Relation to Multi-Stakeholder Pharmaceutical Games}
Our findings connect naturally with the three-party evolutionary game of Feng et al.\ \cite{feng2024}, who model regulators, pharmaceutical companies, and patients in the context of accessible (Braille-labeled) drug packaging under China's legal framework. Both models use replicator dynamics over three boundedly rational populations and evaluate stability through the eigenvalues of the system Jacobian. The structural conclusions are complementary. Feng et al.\ find that administrative penalties, production subsidies, and insurance reimbursement rates are the main drivers of their system's evolution toward a stable state. Our calibrated model shows that the cost side matters just as much: once friction exceeds the combined incentive, $k^3>S_{pol}+\alpha+\gamma-C_{base}$, no face center remains in the domain and no penalty schedule at those levels can hold the market in compliance (\cref{eq:k2star}). Two further differences follow from the structure of our model. Since the calibrated populations face different cost-to-benefit ratios, there is no interior fixed point at all, and the long-run state lies on a face of the cube (\cref{eq:V}); and because the face dynamics conserve $\mathcal{H}$, compliance cycles neutrally rather than settling at an evolutionarily stable state. In Feng et al.'s framing, our threshold result says that the transition to the desirable compliant state is achievable only while regulation remains cheap relative to its incentives; a regulator that becomes expensive to comply with undermines the very behavior it is trying to enforce.

\subsection{Broader Impacts across EGT Sectors}
The framework developed here has implications beyond pharmaceutical governance. The path dependence found in Panel~3.4 suggests a general tendency in markets governed by replicator-type selection: once an industry has nearly abandoned compliance, restoring it within a given horizon requires a somewhat larger incentive than was needed to maintain it, because recovery from close to the boundary is slow.

The same dynamic may apply to emerging sectors such as artificial intelligence (AI) regulation and environmental carbon-tax policy. If the friction of AI safety testing becomes too high, technology firms may shift to defection, accepting government fines as acceptable operational overhead. To address the volunteer's dilemma without relying on direct taxpayer funding, policy design could focus on expanding non-punitive, exogenous state assets such as priority review vouchers (\cref{fig:prv}). In the model such instruments raise $S_{pol}$ and so shift the thresholds $k^*$ and $k^{**}$ upward, which can make compliance the better response at a given level of friction.

\subsection{Limitations}\label{sec:limitations}
Several limitations must be acknowledged. First, the ODEs rely on a mean-field assumption, treating the industry as an infinite, well-mixed population and abstracting away the disproportionate market power of large conglomerates. Second, the composite indices ($x,y,z,k$) are synthesized from proxy variables, such as warning letters, recall rates, and media sentiment, which may not capture the full spectrum of corporate ethics or regulatory friction, and the firm sample is stratified rather than random (\cref{sec:data}). Every firm-level number refers either to fiscal 2019 (R\&D and revenue) or to 1991--2021 (penalties) and comes from a published source listed in Appendix~\ref{app:firms}. Since the sample consists of the most heavily penalized companies, it describes how the burden of penalties differs across strategies among firms with substantial enforcement histories; it cannot say whether low-R\&D firms are more likely to be penalized in the first place.  Third, the calibration has limited power: eight biennial observations are used to fit eight constants, four of which finish on a bound of the optimizer (\cref{tab:params}), and we report a single best fit without parameter uncertainty. The equality $\gamma=C_{base}$ that results from two constants sharing a bound creates the degenerate edge noted in \cref{sec:jac}, and the sensitivity analysis shows that breaking it changes the long-run compliance by about $0.3$ (Panel~4.4). Fourth, the long-run analysis holds $k$ constant, whereas the empirical friction rose throughout 2010--2024; the thresholds describe where the long-run behavior changes, not how fast the transient responds. Relatedly, the model's natural cycle lasts roughly a century in calibrated time units, much longer than the fourteen-year data window, so the data constrain only a fraction of one cycle. Finally, the model holds the payoff constants fixed across the whole timeline, whereas real-world economic conditions and legislative mandates change continuously. These constraints provide a roadmap for future iterations of the framework: time-varying payoffs, uncertainty quantification for the fitted constants, a calibration in which the fitted constants are interior to their bounds, and network topologies in place of the mean-field assumption.

\subsection{Future Directions: Regulatory Heterogeneity and Market Segmentation}
The model is calibrated to the FDA, but the same framework could be re-parameterized for other regulators (e.g., the EMA or Japan's PMDA), which differ in approval delay and post-market recall rates and would therefore imply different friction ($k$) and subsidy ($S_{pol}$) parameters. Future work could examine how these regulatory structures shift the thresholds $k^*$ and $k^{**}$. A second natural extension relaxes the mean-field assumption: because generic, novel biologic, and orphan drug developers face different payoffs, the spatial lattice here could be generalized to agent-based scale-free networks (e.g., a Barab\'asi--Albert graph) in which large lobbying groups act as high-degree ``super-nodes.'' Both directions are left to future work.

\subsection{Future Directions: Time Delays and the Social Media Acceleration}
Our delayed model isolates the FDA's bureaucratic lag. A more comprehensive model would account for delays in all three populations: firms take time to pivot pipelines ($\tau_x$), regulators to prosecute ($\tau_y$), and the public to organize ($\tau_z$). The architecture of public pressure is also changing. Information cascades generated by social media algorithms have the potential to shrink the public reaction delay ($\tau_z \rightarrow 0$) \cite{bollen2011}. As awareness of unethical pricing or side effects spreads faster, $z$ could rise rapidly. Future models could use stochastic differential equations with Poisson jump processes to predict how sudden bursts of public attention alter the survival probability of defecting firms. Since between $k^*$ and $k^{**}$ the model already assigns enforcement to public scrutiny (\cref{fig:verif_bif}), such acceleration could make $\gamma$ the dominant evolutionary enforcer of the next decade.

\section{Conclusion}
By treating the U.S.\ pharmaceutical industry as three interacting populations governed by evolutionary game theory, this analysis suggests that corporate ethical behavior is not a permanent state but a recurring cycle. In the calibrated model there is no interior equilibrium; firms and regulators cycle neutrally around a center whose existence is guaranteed by a conserved quantity, and public scrutiny is gradually selected out while regulators can carry enforcement. Bureaucratic friction organizes two thresholds: above $k^*\approx0.75$ regulators saturate and public scrutiny must supply the remaining incentive, and above $k^{**}\approx0.81$ compliance collapses toward a state in which regulators and the public are fully mobilized but no firm complies. The empirical friction index crossed both thresholds between 2016 and 2020, the period of the observed compliance decline, and the 2024 indices lie near the predicted collapse state. Poorly calibrated bureaucracy can therefore degrade healthcare outcomes by pushing the industry past these thresholds. In the absence of incentives that scale with friction, such as state subsidies and an active, punitive public, the model's selection pressures work against corporate ethics; sustaining positive public-health outcomes requires policy that keeps friction below the combined incentive to comply.

To make these trade-offs explorable rather than merely asserted, we also built an interactive dashboard over the calibrated system, in which bureaucratic friction $k$, the state subsidy $S_{pol}$, and the public penalty $\gamma$ can be varied while the population trajectories are recomputed in real time. It is documented in Appendix~\ref{app:dashboard}; we found it useful for locating the collapse threshold by hand before confirming it analytically.

\section*{Author Contributions}
Contributions follow the CRediT taxonomy. \textbf{Alanna Polyak:} Conceptualization; Methodology; Software; Formal analysis; Data curation; Visualization; Writing -- original draft; Writing -- review \& editing. \textbf{Thomas Zdyrski:} Writing -- review \& editing; Supervision; Project administration.

\section*{Acknowledgements}
We would like to acknowledge Denny Bao, Brooke Hollawell, Annika Jiwani, and Ekam Singh for their contributions during the conceptualization, data gathering, and first draft phases of the project. We also like to thank Thomas Kornfeld for his contributions to reviewing aspects of the formal analysis phase of this project.

\section*{Data and Code Availability}
The code repository is available on GitHub at \url{https://github.com/alannap27/EGT-Pharma-Ethics}
and Zenodo at \url{https://doi.org/10.5281/zenodo.23605707}.
It is organized with a top-level \texttt{README}; the model module \texttt{egt\_core.py}, which implements the calibration and every analytic quantity used in the paper; the notebook \texttt{egt\_pharma\_analysis.ipynb} which regenerates every figure and also runs stand-alone; and the data folder with the composite indices and the firm-level dataset of Appendix~\ref{app:firms}, including the source document and access date for every penalty and revenue figure.
The interactive notebook is also available on \href{https://mybinder.org/v2/gh/alannap27/EGT-Pharma-Ethics/HEAD?urlpath=%2Fdoc%2Ftree%2Finteractive_dashboard.ipynb}{MyBinder}.

\section*{Declaration of Interests}
The authors declare that they have no known competing financial interests or personal relationships that could have appeared to influence the work reported in this paper.

\section*{Disclosure of AI Use}
Generative AI tools were used to assist with editing portions of the code and manuscript text as well as with comment formatting in the code; AI was not used to solely generate any of the mathematical findings or to wholly produce any figure. All models, analyses, results, and conclusions were reviewed and verified by the authors.

\printbibliography

\begin{appendices}
\crefalias{section}{appendix}
\section{Firm Data and Strategy Classification}\label{app:firms}
Since no single public database tracks evolutionary game-theory metrics for pharmaceutical behavior, we manually compiled the dataset below. Using fiscal-2019 annual reports, Form 10-K filings, and the Public Citizen penalty tabulation, we categorized 15 major firms into three strategy groups:
\begin{itemize}
    \item \textbf{S1 -- ethical R\&D (cooperators):} spend heavily on foundational research ($>21\%$ of revenue).
    \item \textbf{S2 -- cost cutters (defectors):} spend little on internal R\&D ($<10\%$ of revenue), drawing revenue from acquired or repriced legacy drugs, and carry the largest penalties relative to revenue (\cref{fig:zonehist}).
    \item \textbf{Mixed strategy:} R\&D intensity of $10$--$21\%$ of revenue; solid pipelines but recurrent settlements, mostly for off-label marketing.
\end{itemize}

\paragraph{Firm selection:} The 15 firms are every company on Public Citizen's top-20 list of penalty totals \cite{publiccitizen2024} that still existed as an independent company in fiscal 2019 and for which fiscal-2019 R\&D expense, fiscal-2019 revenue, and the penalty total are all published (\cref{sec:data}). Purdue Pharma and Reckitt Benckiser, which are on the list, are excluded for lack of published R\&D figures; Schering-Plough, TAP, and Serono no longer existed in 2019.

\paragraph{Sources:} Penalties are each firm's total federal and state financial penalties over 1991--2021 from Table~4 of Public Citizen \cite{publiccitizen2024}; principal settlement amounts in the table are from the same report. R\&D intensity is fiscal-2019 R\&D expense divided by fiscal-2019 revenue, and revenue is fiscal-2019 revenue, both as reported in each firm's annual report, Form 10-K, or Form 20-F (for Pfizer, as originally reported in the fiscal-2019 Form 10-K; GlaxoSmithKline's turnover is converted to U.S. dollars). The document, figures, and access date for every number are recorded in \texttt{data/firm\_classification.csv}.

{\footnotesize
\begin{xltabular}{\textwidth}{@{} >{\raggedright\arraybackslash}p{2.3cm} >{\raggedright\arraybackslash}p{2.2cm} L >{\raggedright\arraybackslash}p{1.4cm} L @{}}
\caption{The fifteen firms, by strategy zone. Penalties are government financial penalties over 1991--2021 \cite{publiccitizen2024}, as plotted in \cref{fig:rnd_penalty}; R\&D intensity and revenue are for fiscal 2019 (sources in the text above).}\label{tab:full_firms}\\
\toprule
\textbf{Company} & \textbf{Strategy} & \textbf{Principal violations; penalties 1991--2021} & \textbf{R\&D \%} & \textbf{Revenue 2019} \\
\midrule
\endfirsthead
\multicolumn{5}{@{}l}{\textit{Table~\thetable\ (continued)}}\\
\toprule
\textbf{Company} & \textbf{Strategy} & \textbf{Principal violations; penalties 1991--2021} & \textbf{R\&D \%} & \textbf{Revenue 2019} \\
\midrule
\endhead
\bottomrule
\endlastfoot
Teva & S2 -- Defect & Monopoly practices (\$1.2B, 2015), among 22 settlements. Total \$2.33B. & 6.0\% & \$16.89B \\ \addlinespace
Abbott & S2 -- Defect & Depakote off-label promotion (\$1.5B, 2012), among 17 settlements. Total \$2.01B. & 7.6\% & \$31.90B \\ \addlinespace
Mylan (Viatris) & S2 -- Defect & EpiPen misclassified to underpay Medicaid rebates (\$465M, 2017). Total \$1.21B. & 5.6\% & \$11.50B \\ \addlinespace
Indivior & S2 -- Defect & Suboxone marketing (\$600M federal resolution, 2020). Total \$1.18B. & 6.8\% & \$0.79B \\ \addlinespace
Mallinckrodt & Mixed & Opioid marketing, multistate (\$1.6B, 2020); Medicaid underpayment. Total \$1.75B. & 11.0\% & \$3.16B \\ \addlinespace
Allergan & Mixed & Four settlements. Total \$0.82B. & 11.3\% & \$16.09B \\ \addlinespace
Glaxo\-Smith\-Kline & Mixed & Paxil, Wellbutrin and Avandia promotion and withheld data (\$3.0B, 2012). Total \$7.92B. & 13.5\% & \$43.92B \\ \addlinespace
Johnson \& Johnson & Mixed & Risperdal off-label marketing (\$2.0B, 2013); opioid multistate (\$5.0B, 2021). Total \$8.45B. & 13.8\% & \$82.06B \\ \addlinespace
Pfizer & Mixed & Off-label promotion of Bextra and other drugs (\$2.3B, 2009). Total \$4.77B. & 16.7\% & \$51.75B \\ \addlinespace
Amgen & Mixed & Aranesp, Enbrel and Neulasta promotion and kickbacks (\$762M, 2012). Total \$0.94B. & 17.6\% & \$23.36B \\ \addlinespace
Novartis & Mixed & Kickbacks (\$678M, 2020), among 25 settlements. Total \$2.95B. & 19.8\% & \$47.45B \\ \addlinespace
Merck & S1 -- Ethical & Vioxx unlawful promotion (\$950M, 2011). Total \$1.84B. & 21.1\% & \$46.84B \\ \addlinespace
Bristol Myers Squibb & S1 -- Ethical & Plavix judgment with Sanofi (\$834M, 2021), among 16 settlements. Total \$1.82B. & 23.5\% & \$26.15B \\ \addlinespace
AstraZeneca & S1 -- Ethical & Seroquel off-label marketing, among 13 settlements. Total \$1.04B. & 24.8\% & \$24.38B \\ \addlinespace
Eli Lilly & S1 -- Ethical & Off-label Zyprexa promotion (\$1.415B, 2009). Total \$1.74B. & 25.1\% & \$22.32B \\
\end{xltabular}}

\section{Parameter Values}\label{app:params}
\Cref{tab:params} lists every constant used in the paper, \cref{tab:derived} the quantities derived from them, and \cref{tab:indices} the composite indices to which the model is fitted.

\begin{table}[H]
\centering
\caption{Payoff constants fitted by bounded L-BFGS-B (\cref{sec:ml}), with the bounds used, and the fixed constants used in the extensions. ``Bound'' marks a fitted value that finishes on a bound. The optimizer starts from $(C_{base},S_{pol},\alpha,B_r,C_r,\gamma,B_p,C_p)=(0.2,0.4,0.2,0.8,0.25,0.6,0.6,0.15)$.}\label{tab:params}
{\small
\begin{tabular}{@{}llccl@{}}
\toprule
Symbol & Meaning & Value & Bounds & Note \\
\midrule
$C_{base}$ & baseline cost of compliance & $0.1000$ & $[0.1,0.5]$ & bound \\
$S_{pol}$ & FDA subsidy to complying firms & $0.3209$ & $[0.1,1.0]$ & \\
$\alpha$ & FDA fine on defectors & $0.2078$ & $[0.1,1.0]$ & \\
$B_r$ & regulator benefit from catching a defector & $1.0000$ & $[0.5,1.0]$ & bound \\
$C_r$ & regulator overhead & $0.1000$ & $[0.1,0.5]$ & bound \\
$\gamma$ & public penalty on defectors & $0.1000$ & $[0.1,1.0]$ & bound \\
$B_p$ & public benefit from catching a defector & $0.8977$ & $[0.4,1.0]$ & \\
$C_p$ & public overhead & $0.1017$ & $[0.1,0.5]$ & \\
\midrule
$\sigma$ & noise amplitude, Panel 3.1 & $0.08$ & --- & illustrative \\
--- & log-normal shock scale, Panel 3.3 & $0.5$ & --- & illustrative \\
$\tau_y$ & regulator delay, Panel 3.5 & $1.5$ yr & --- & illustrative \\
$b$ & temptation payoff, \cref{fig:spatial} & $1.65$ & --- & illustrative \\
--- & objective weights & $8\times$ on $x$; time ramp $1\to3$ & --- & \cref{sec:ml} \\
--- & L-BFGS-B starting point & see note & --- & \\
\bottomrule
\end{tabular}}
\end{table}

\begin{table}[H]
\centering
\caption{Quantities derived from the fitted constants.}\label{tab:derived}
{\small
\begin{tabular}{@{}lll@{}}
\toprule
Quantity & Expression & Value \\
\midrule
$x$-coordinate of $E_z$, $E_{z1}$ & $x_z^*=1-C_r/B_r$ & $0.900$ \\
$x$-coordinate of $E_y$, $E_{y0}$ & $x_y^*=1-C_p/B_p$ & $0.887$ \\
drift of $V$ (\cref{eq:V}) & $x_z^*-x_y^*$ & $0.0133$ \\
transverse decay rate of $z$ & $B_p(x_y^*-x_z^*)$ & $-0.0119$ \\
threshold where $E_z$ reaches $y=1$ & $k^*=(S_{pol}+\alpha-C_{base})^{1/3}$ & $0.754$ \\
threshold where $E_y$ reaches $z=1$ & $k^{**}=(S_{pol}+\alpha+\gamma-C_{base})^{1/3}$ & $0.809$ \\
$E_z$ at $k=0.60$ & $(x_z^*,\,C_{base}/(S_{pol}+\alpha-k^3),\,0)$ & $(0.900,\,0.320,\,0)$ \\
small-orbit period at $k=0.60$ & $2\pi/\omega$ & $80.3$ \\
$E_y$ at $k=\kbar$ & $(x_y^*,\,1,\,(C_{base}-S_{pol}-\alpha+k^3)/\gamma)$ & $(0.887,\,1,\,0.785)$ \\
small-orbit period at $k=\kbar$ & $2\pi/\omega$ & $161$ \\
historical mean friction & $\kbar$ & $0.7975$ \\
\bottomrule
\end{tabular}}
\end{table}

\begin{table}[H]
\centering
\caption{Composite indices (biennial, 2010--2024) used for calibration; one time unit is one year.}\label{tab:indices}
{\small
\begin{tabular}{@{}lcccccccc@{}}
\toprule
Year & 2010 & 2012 & 2014 & 2016 & 2018 & 2020 & 2022 & 2024 \\
\midrule
ethical firms $x$ & 0.65 & 0.72 & 0.78 & 0.85 & 0.80 & 0.60 & 0.40 & 0.25 \\
regulators $y$ & 0.20 & 0.30 & 0.45 & 0.60 & 0.75 & 0.85 & 0.92 & 0.88 \\
public $z$ & 0.15 & 0.18 & 0.25 & 0.35 & 0.50 & 0.65 & 0.80 & 0.90 \\
friction $k$ & 0.60 & 0.65 & 0.70 & 0.75 & 0.80 & 0.88 & 0.95 & 1.05 \\
\bottomrule
\end{tabular}}
\end{table}

\section{Conserved Quantity and Centers}\label{app:hamiltonian}
This appendix gives the derivation behind \cref{eq:H}. It is expository rather than a formal proof. We work on the invariant face $z=0$; the faces $y=1$, $y=0$, and $z=1$ follow by the substitutions noted at the end.

\paragraph{Enumeration of the fixed points:} Fixed points require $\dot x=\dot y=\dot z=0$. Each equation in \cref{eq:generic} vanishes if its coordinate is $0$ or $1$ or if its bracket vanishes, which gives the $3\times3\times3=27$ combinations of \cref{tab:27}; a fixed point is obtained by choosing one column, and one of the three options in that column's bottom row.

\begin{table}[H]
\centering
\caption{The 27 combinations that make $\dot z$, $\dot y$, and $\dot x$ vanish. Each column fixes how $\dot z=0$ and $\dot y=0$ are satisfied; the bottom row lists the ways in which $\dot x=0$ can then hold.}\label{tab:27}
\renewcommand{\arraystretch}{1.25}
\resizebox{\textwidth}{!}{%
\begin{tabular}{|r|c|c|c|c|c|c|c|c|c|}
\hline
$\dot{z} = 0$: & \multicolumn{3}{c|}{$z = 0$} & \multicolumn{3}{c|}{$z = 1$} & \multicolumn{3}{c|}{$x = -c_0/c_x$} \\ \hline
$\dot{y} = 0$: & $y=0$ & $y=1$ & $x = -b_0/b_x$ & $y=0$ & $y=1$ & $x = -b_0/b_x$ & $y=0$ & $y=1$ & $b_0/b_x = c_0/c_x$ \\ \hline
$\dot{x} = 0$: &
$\begin{aligned} x&=0 \\ \text{or}\;x&=1 \\ \text{or}\;a_0&=0 \end{aligned}$ &
$\begin{aligned} x&=0 \\ \text{or}\;x&=1 \\ \text{or}\;a_0&=-a_y \end{aligned}$ &
$\begin{aligned} b_0&=0 \\ \text{or}\;b_0&=-b_x \\ \text{or}\;y&=-a_0/a_y \end{aligned}$ &
$\begin{aligned} x&=0 \\ \text{or}\;x&=1 \\ \text{or}\;a_0&=-a_z \end{aligned}$ &
$\begin{aligned} x&=0 \\ \text{or}\;x&=1 \\ \text{or}\;a_0&=-a_y-a_z \end{aligned}$ &
$\begin{aligned} b_0&=0 \\ \text{or}\;b_0&=-b_x \\ \text{or}\;y&=-\tfrac{a_0+a_z}{a_y} \end{aligned}$ &
$\begin{aligned} c_0&=0 \\ \text{or}\;c_0&=-c_x \\ \text{or}\;z&=-a_0/a_z \end{aligned}$ &
$\begin{aligned} c_0&=0 \\ \text{or}\;c_0&=-c_x \\ \text{or}\;z&=-\tfrac{a_0+a_y}{a_z} \end{aligned}$ &
$\begin{aligned} c_0&=0 \\ \text{or}\;c_0&=-c_x \\ \text{or}\;z&=-\tfrac{a_0+a_y y}{a_z} \end{aligned}$ \\ \hline
\end{tabular}}
\end{table}
Of the 27 combinations, 8 are the corners of the unit cube ($x,y,z\in\{0,1\}$). Fifteen exist only for special values of the constants (for example $a_0=0$ or $b_0=-b_x$); of these the only one of interest is $x=-c_0/c_x=-b_0/b_x$, which gives the line of fixed points $z=-(a_0+a_y y)/a_z$. The 4 remaining combinations are the face fixed points of \cref{eq:facefp},
\begin{equation*}
E_{y0}=\Big(-\tfrac{c_0}{c_x},\,0,\,-\tfrac{a_0}{a_z}\Big),\quad
E_y=\Big(-\tfrac{c_0}{c_x},\,1,\,-\tfrac{a_0+a_y}{a_z}\Big),\quad
E_z=\Big(-\tfrac{b_0}{b_x},\,-\tfrac{a_0}{a_y},\,0\Big),\quad
E_{z1}=\Big(-\tfrac{b_0}{b_x},\,-\tfrac{a_0+a_z}{a_y},\,1\Big).
\end{equation*}
The points on the faces $y=0$ and $y=1$ share the $x$-coordinate $x_y^*=-c_0/c_x$, and those on the faces $z=0$ and $z=1$ share $x_z^*=-b_0/b_x$.

\paragraph{Stability of the corners:} At a corner every off-diagonal entry of \cref{eq:jacobian} contains a factor $x(1-x)$, $y(1-y)$, or $z(1-z)$ and vanishes, so
\begin{equation*}
J\big|_{\text{corner}}=\operatorname{diag}\big((1-2x)F,\;(1-2y)G,\;(1-2z)H\big),
\end{equation*}
with eigenvalues $\pm F,\pm G,\pm H$, the sign depending on whether each coordinate is $0$ or $1$. These are real and generically nonzero, so the corners are hyperbolic fixed points of node or saddle type.

\paragraph{Stability of the face fixed points:} Evaluating \cref{eq:jacobian} at each face point, where the corresponding brackets vanish, gives
\begin{align*}
J\big|_{E_{y0}} &=
\begin{pmatrix}
0 & -\frac{(c_0+c_x)c_0 a_y}{c_x^2} & -\frac{(c_0+c_x)c_0 a_z}{c_x^2} \\
0 & \frac{b_0c_x - b_x c_0}{c_x} & 0 \\
-\frac{(a_0+a_z)a_0 c_x}{a_z^2} & 0 & 0
\end{pmatrix},\\
&\qquad \lambda = \frac{b_0c_x - b_x c_0}{c_x},\;\; \pm\sqrt{\frac{(a_0+a_z)(c_0+c_x)\,a_0 c_0}{a_z c_x}},\\[6pt]
J\big|_{E_{y}} &=
\begin{pmatrix}
0 & -\frac{(c_0+c_x)c_0 a_y}{c_x^2} & -\frac{(c_0+c_x)c_0 a_z}{c_x^2} \\
0 & -\frac{b_0c_x - b_x c_0}{c_x} & 0 \\
-\frac{(a_0+a_y+a_z)(a_0+a_y) c_x}{a_z^2} & 0 & 0
\end{pmatrix},\\
&\qquad \lambda = -\frac{b_0c_x - b_x c_0}{c_x},\;\; \pm\sqrt{\frac{(a_0+a_y+a_z)(c_0+c_x)(a_0+a_y)\,c_0}{a_z c_x}},\\[6pt]
J\big|_{E_{z}} &=
\begin{pmatrix}
0 & -\frac{(b_0+b_x)b_0 a_y}{b_x^2} & -\frac{(b_0+b_x)b_0 a_z}{b_x^2} \\
-\frac{(a_0+a_y)a_0 b_x}{a_y^2} & 0 & 0 \\
0 & 0 & \frac{c_0b_x - c_x b_0}{b_x}
\end{pmatrix},\\
&\qquad \lambda = \frac{c_0b_x - c_x b_0}{b_x},\;\; \pm\sqrt{\frac{(a_0+a_y)(b_0+b_x)\,a_0 b_0}{a_y b_x}},\\[6pt]
J\big|_{E_{z1}} &=
\begin{pmatrix}
0 & -\frac{(b_0+b_x)b_0 a_y}{b_x^2} & -\frac{(b_0+b_x)b_0 a_z}{b_x^2} \\
-\frac{(a_0+a_y+a_z)(a_0+a_z) b_x}{a_y^2} & 0 & 0 \\
0 & 0 & -\frac{c_0b_x - c_x b_0}{b_x}
\end{pmatrix},\\
&\qquad \lambda = -\frac{c_0b_x - c_x b_0}{b_x},\;\; \pm\sqrt{\frac{(a_0+a_y+a_z)(b_0+b_x)(a_0+a_z)\,b_0}{a_y b_x}}.
\end{align*}
In each case one eigenvalue is transverse to the face and the pair $\pm\sqrt{\cdot}$ has eigenvectors in the face. The $y=1$ case is the $y=0$ case with $a_0\to a_0+a_y$; the $z=0$ case is the $y=0$ case with $b_0\leftrightarrow c_0$, $b_x\leftrightarrow c_x$, and $a_y\leftrightarrow a_z$; and the $z=1$ case is the $y=1$ case under the same exchange. If the expression under the radical is positive, all three eigenvalues are real and the point is a saddle; if it is negative, the in-face pair is purely imaginary and the linearization predicts a center within the face. For $E_z$, substituting $x_z^*=-b_0/b_x$ and $y^*=-a_0/a_y$ shows that $\lambda_\perp=H(x_z^*)=B_p(x_y^*-x_z^*)$ and that the radicand equals $\lambda_\parallel^2$ of \cref{eq:eigEz}.

\paragraph{A conserved quantity:} On $z=0$ the system reduces to $\dot x=x(1-x)(a_0+a_y y)$, $\dot y=y(1-y)(b_0+b_x x)$. We seek $\mathcal{H}(x,y)$ with $\dot{\mathcal{H}}=\partial_x\mathcal{H}\,\dot x+\partial_y\mathcal{H}\,\dot y=0$, i.e.,
\begin{equation*}
x(1-x)(a_0+a_y y)\,\frac{\partial \mathcal{H}}{\partial x} + y(1-y)(b_0+b_x x)\,\frac{\partial \mathcal{H}}{\partial y} = 0 .
\end{equation*}
Assuming a separable solution $\mathcal{H}=X(x)\,Y(y)$ gives
\begin{equation*}
\frac{x(1-x)\,X'}{X\,(b_0+b_x x)} \;=\; -\frac{y(1-y)\,Y'}{Y\,(a_0+a_y y)} .
\end{equation*}
The left side depends only on $x$ and the right only on $y$, so both equal a constant, which we take to be $1$. Integrating by partial fractions,
\begin{align*}
\ln X&=\int\frac{b_0+b_x x}{x(1-x)}\,\mathrm{d} x = b_0\ln x-(b_0+b_x)\ln(1-x),\\
\ln Y&=-\int\frac{a_0+a_y y}{y(1-y)}\,\mathrm{d} y = -a_0\ln y+(a_0+a_y)\ln(1-y),
\end{align*}
which gives \cref{eq:H}. Direct differentiation confirms $\tfrac{\mathrm{d}}{\mathrm{d}t}\ln\mathcal{H}=(b_0+b_x x)F-(a_0+a_y y)G=0$. Note the sign convention $a_0=-C_{base}$, which makes $a_0<0$.

\paragraph{Closed orbits:} When $0<x_z^*<1$ and $0<y^*<1$, i.e., $b_0>0>b_0+b_x$ and $a_0<0<a_0+a_y$, each of the four terms of $\ln\mathcal{H}$ tends to $-\infty$ on one edge of the face, and $\ln\mathcal{H}$ is strictly concave in each variable with its unique stationary point at $E_z$. Every level set of $\mathcal{H}$ inside the face is therefore a closed curve around $E_z$. Since trajectories stay on level sets, every orbit on the face is periodic, and $E_z$ is a true center: an isolated fixed point at a strict extremum of a conserved quantity is surrounded by closed orbits \cite[Sec.~6.5]{strogatz2024}. The time average of $x$ over any of these orbits is $x_z^*$, and that of $y$ is $y^*$, because $\ln[y/(1-y)]$ and $\ln[x/(1-x)]$ return to their starting values after one period.

\paragraph{Hamiltonian structure:} The face system is not Hamiltonian in $(x,y)$: its divergence $\partial_x\dot x+\partial_y\dot y=(1-2x)F+(1-2y)G$ does not vanish, so area is not conserved. It is, however, conformally Hamiltonian \cite{garcia2020}: one checks directly that
\begin{equation*}
\dot x = \phi(x,y)\,\frac{\partial \mathcal{H}}{\partial y},\qquad \dot y=-\phi(x,y)\,\frac{\partial \mathcal{H}}{\partial x},\qquad \phi(x,y)=-\frac{x(1-x)\,y(1-y)}{\mathcal{H}},
\end{equation*}
so after the state-dependent rescaling of time $\mathrm{d}\tau=\phi\,\mathrm{d} t$ the system is Hamiltonian with Hamiltonian $\mathcal{H}$. Equivalently, in log-odds coordinates $u=\ln[x/(1-x)]$, $v=\ln[y/(1-y)]$ the system reads $\dot u=a_0+a_y y(v)$, $\dot v=b_0+b_x x(u)$, which is exactly Hamiltonian, $\dot u=\partial K/\partial v$ and $\dot v=-\partial K/\partial u$, with
\begin{equation*}
K(u,v) = a_0 v + a_y\ln(1+e^{v}) - b_0 u - b_x\ln(1+e^{u}) = -\ln\mathcal{H},
\end{equation*}
a known property of the replicator dynamics of $2\times2$ bimatrix games \cite{hofbauer1996}.

\paragraph{Other faces:} On the face $y=1$, the same derivation with $(a_0,a_y,b_0,b_x)\to(a_0+a_y,a_z,c_0,c_x)$ gives the conserved quantity $\mathcal{H}_y(x,z)=x^{c_0}(1-z)^{a_0+a_y+a_z}/\big[(1-x)^{c_0+c_x}z^{a_0+a_y}\big]$ and shows that $E_y$ is a true center whenever it lies in the face. The faces $y=0$ and $z=1$ follow analogously; under the fitted constants their fixed points lie on the degenerate edge $y=0$, $z=1$.

\section{Interactive Replicator Dynamics Dashboard}\label{app:dashboard}
\begin{figure}[H]
    \centering
    \includegraphics[width=\textwidth]{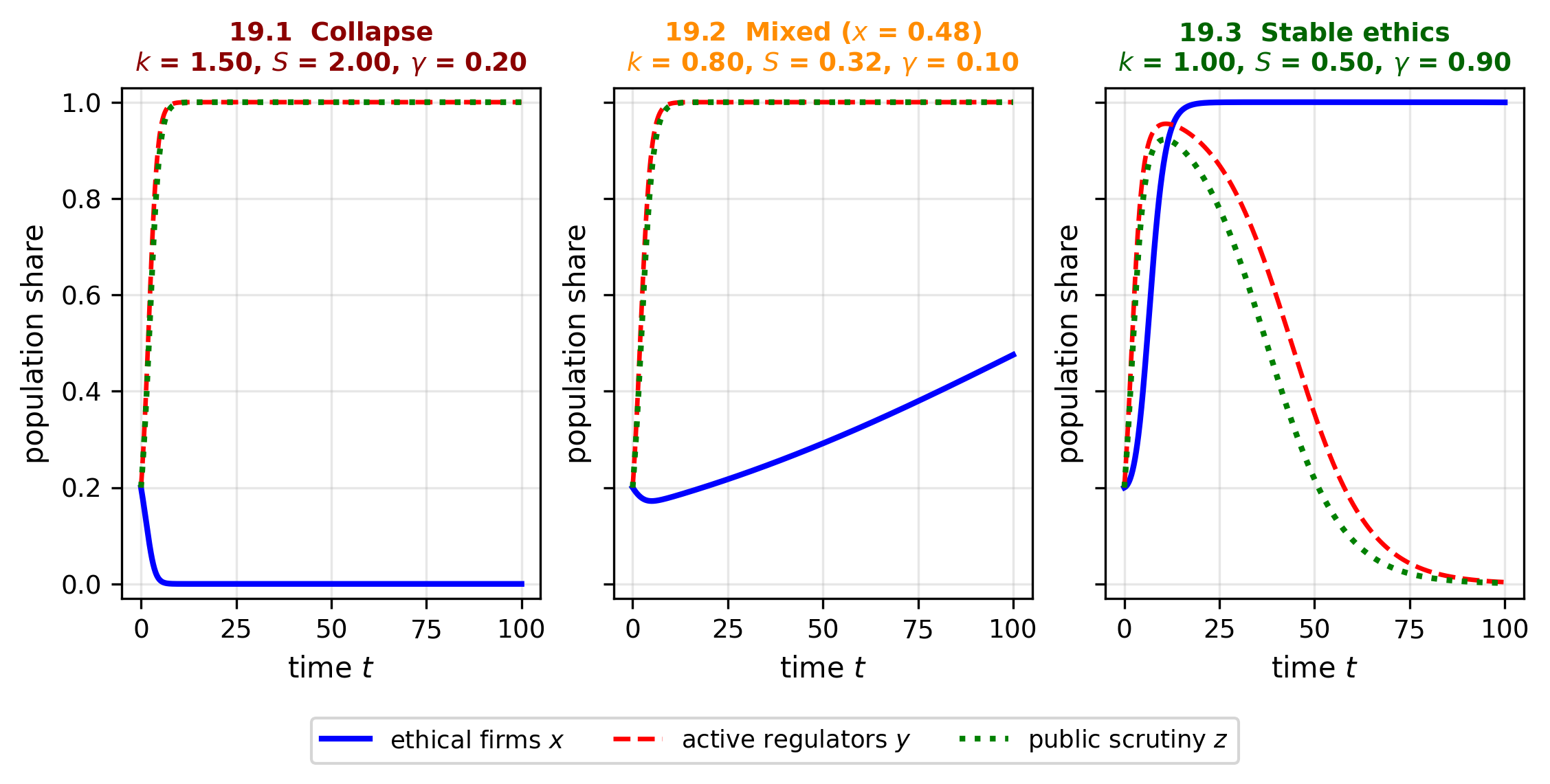}
    \caption{\textbf{Dashboard examples}, each from $(0.2,0.2,0.2)$ over $0\le t\le100$ with the other constants at their fitted values. (19.1) collapse: $(k,S_{pol},\gamma)=(1.50,2.00,0.20)$; (19.2) mixed: $(0.80,0.32,0.10)$, close to the fitted subsidy and penalty at the historical friction; (19.3) stable ethics: $(1.00,0.50,0.90)$.}\label{fig:dashboard}
\end{figure}

To allow parameter stress-testing, the \href{https://github.com/alannap27/EGT-Pharma-Ethics}{GitHub repository} includes a \texttt{interactive\_dashboard.ipynb} notebook with an interactive dashboard built with the \texttt{ipywidgets} framework. It exposes three sliders, for bureaucratic friction $k$, the FDA subsidy $S_{pol}$, and the public penalty $\gamma$, holds the remaining constants at their fitted values, and recomputes the trajectory from $(0.2,0.2,0.2)$ over $0\le t\le100$ whenever a slider moves. When \texttt{ipywidgets} is unavailable, the \texttt{egt\_pharma\_analysis.ipynb} and \texttt{interactive\_dashboard.ipynb} notebooks render the three example settings of \cref{fig:dashboard} statically.
An interactive version is hosted on \href{https://mybinder.org/v2/gh/alannap27/EGT-Pharma-Ethics/HEAD?urlpath=%2Fdoc%2Ftree%2Finteractive_dashboard.ipynb}{MyBinder} for easier inspection.

\textbf{Analysis of the dashboard examples:} \Cref{fig:dashboard} shows the three outcome classes the dashboard reports. It labels a run ``collapse'' if compliance at $t=100$ is at most $0.15$ and ``stable ethics'' if it is at least $0.85$. In \textbf{Panel 19.1}, extreme friction overwhelms even a large subsidy ($S_{pol}+\alpha+\gamma-C_{base}=2.31<k^3=3.38$), and compliance collapses while regulators and the public saturate. In \textbf{Panel 19.2}, the system sits just above the threshold ($k^{**}=0.808$ for these values): compliance recovers, but only slowly, reaching $0.48$ at $t=100$. Lowering the subsidy to $0.30$ moves $k^{**}$ below $0.80$ and turns this example into a collapse, which illustrates how sharply the outcome depends on the threshold. In \textbf{Panel 19.3}, a large public penalty restores full compliance, after which regulators and the public withdraw because they are no longer needed.

\end{appendices}

\end{document}